\documentclass[10pt,journal,compsoc]{IEEEtran}

\usepackage{xcolor}

\definecolor{teal}{RGB}{0,128,128}

\newcommand{\citet}[1]{\cite{#1}}

\newtheorem{definition}{Definition}

\usepackage{microtype}

\usepackage{mathtools}
\DeclarePairedDelimiterX\setc[2]{\{}{\}}{\,#1 \;\delimsize\vert\;#2\,}
\usepackage{xspace}
\newcommand{\algo}{\textit{merge}\xspace}

\newcommand{\naive}{{\it na\"ive}}
\newcommand{\ltre}{{\it LTRE}}

\newcommand{\deepMergeTool}{{\sc DeepMerge}}
\newcommand{\tool}{\deepMergeTool}
\newcommand{\mergetomatrix}{\textit{Merge2Matrix}\xspace}

\usepackage{subcaption}
\usepackage{listings}
\usepackage{mathtools}

\usepackage{amssymb}
\usepackage{enumitem}
\usepackage{outlines}
\usepackage{tikz}
\usepackage{multirow}
\usepackage{amsmath}
\usepackage{comment}
\DeclareMathOperator*{\argmax}{arg\,max}

\usepackage{algpseudocode} 
\usepackage{algorithm}
\usepackage{booktabs}

\makeatletter
\newcommand{\linebreakand}{%
  \end{@IEEEauthorhalign}
  \hfill\mbox{}\par
  \mbox{}\hfill\begin{@IEEEauthorhalign}
}
\makeatother

\usepackage[english]{babel}
\usepackage{blindtext}
\input{code_highlight.tex}

\begin{document}

%\title{DeepMerge: Language Agnostic Merge Conflict Resolution as a Deep Learning Task}
% \title{DeepMerge: Merge Conflict Resolution as a Machine Learning Task}
% \title{DeepMerge: Data-Driven Three-Way Program Merge}
%\title{DeepMerge: Learning to Perform Three-Way Program Merge}
\title{DeepMerge: Learning to Merge Programs}%\thanks{Preliminary versions of this paper were made available on arXiv at  https://arxiv.org/abs/2105.07569.  This version of the paper has been substantially improved based on feedback of the earlier versions.}}

\author{Elizabeth~Dinella,~\IEEEmembership{Member,~IEEE,}
  Todd~Mytkowicz,~\IEEEmembership{Member,~IEEE,}
  Alexey~Svyatkovskiy,~\IEEEmembership{Member,~IEEE}
  Christian~Bird,~\IEEEmembership{Distinguished Scientist,~IEEE}
  Mayur~Naik,~\IEEEmembership{Member,~IEEE}
  Shuvendu~Lahiri,~\IEEEmembership{Member,~IEEE}% <-this % stops a space
\IEEEcompsocitemizethanks{\IEEEcompsocthanksitem Elizabeth Dinella is with the University of Pennsylvania, Philadelphia, Pennsylvania, USA. E-mail: edinlla@seas.upenn.edu. Elizabeth Dinella performed this work while employed at Microsoft Research. 
\IEEEcompsocthanksitem Todd~Mytkowicz is with Microsoft Research, Redmond, Washington, USA. E-mail: toddm@microsoft.com
\IEEEcompsocthanksitem Alexey~Svyatkovskiy is with Microsoft, Redmond, Washington, USA. E-mail: alsvyatk@microsoft.com
\IEEEcompsocthanksitem Christian Bird is with Microsoft Research, Redmond, Washington, USA. E-mail: cbird@microsoft.com
\IEEEcompsocthanksitem Mayur Naik is with the University of Pennsylvania, Philadelphia, Pennsylvania, USA. E-mail: mhnaik@cis.upenn.edu
\IEEEcompsocthanksitem Shuvendu Lahiri is with Microsoft Research, Redmond, Washington, USA. E-mail: shuvendu@microsoft.com}% <-this % stops an unwanted space}
}

\maketitle

\begin{abstract}
In collaborative software development, program merging is {\it the} mechanism to integrate changes from multiple programmers. %in modern version control systems. 
Merge algorithms in modern version control systems report a conflict when changes interfere textually. 
Merge conflicts require manual intervention and frequently stall modern continuous integration pipelines. 
%Although costly, it is typically tedious to resolve merge conflicts \shuvendu{Is this a challenge or an opportunity?}. 
Prior work found that, although costly, a large majority of resolutions involve re-arranging text without writing any new code.
Inspired by this observation we propose the {\it first data-driven approach} to resolve merge conflicts with a machine learning model. 
We realize our approach in a tool \deepMergeTool{} that uses a novel combination of 
(i) an edit-aware embedding of merge inputs and
(ii) a variation of pointer networks, to construct resolutions from input segments.
We also propose an algorithm to localize manual resolutions in a resolved file and employ it to curate a ground-truth dataset comprising 8,719 non-trivial resolutions in JavaScript programs.
Our evaluation shows that, on a held out test set, \deepMergeTool{} can predict correct resolutions for 37\% of non-trivial merges, compared to only 4\% by a state-of-the-art semistructured merge technique.
Furthermore, on the subset of merges with upto 3 lines (comprising 24\% of the total dataset), \deepMergeTool{} can predict correct resolutions with 78\% accuracy. 
 
%with high precision ($72$\%) and modest recall ($34$\%) on the dataset overall, 
%and high recall ($78$\%) on merges comprising of upto 3 lines that comprise $24$\%
%of the dataset.

\end{abstract}

% !TeX root = ./main.tex
\newcommand{\baseProg}{\mathcal O}
\newcommand{\aProg}{\mathcal A}
\newcommand{\bProg}{\mathcal B}
\newcommand{\resProg}{\mathcal M}
\newcommand{\mergeProg}{\mathcal C}
\newcommand{\edit}[1]{\Delta_{#1}}
\newcommand{\aEdit}{\edit{\aProg}}
\newcommand{\bEdit}{\edit{\bProg}}
\newcommand{\resEdit}{\edit{\resProg}}

\newcommand{\diffThree}{{\tt diff3}}
\newcommand{\safeMergeTool}{{\sc SafeMerge}}

\newcommand{\var}[1]{{\tt #1}}
\newcommand{\challenge}[1]{{\bf CH{#1}}}

\section{Introduction}
\label{sec:introduction}

In collaborative software development settings, version control systems such as ``git'' are commonplace. 
Such version control systems allow developers to simultaneously edit code through features called branches. 
Branches are a growing trend in version control as they allow developers to work in their own isolated workspace, making changes independently, and only integrating their work into the main line of development when it is complete.
Integrating these changes frequently involves merging multiple copies of the source code. 
In fact, according to a large-scale empirical study of Java projects on GitHub~\cite{ghiotto-tse18}, nearly 12\% of all commits are related to a merge.

To integrate changes by multiple developers across branches, version control systems utilize merge algorithms. Textual three-way file merge (e.g. present in {\it ``git merge''}) is 
the prevailing merge algorithm.
As the name suggests, three-way merge takes three files as input: the common {\it base} file $\baseProg$, and its corresponding modified files, $\aProg$ and $\bProg$. The algorithm either:
\begin{enumerate}
\item declares a ``conflict'' if the two changes interfere with each other, or
\item provides a merged file $\resProg$ that incorporates changes made in $\aProg$ and $\bProg$.
\end{enumerate}

Under the hood, three-way merge typically employs the \diffThree{} algorithm, which performs an {\it unstructured} (line-based) merge~\cite{smith-98}. 
Intuitively, the algorithm {\it aligns} the two-way diffs of $\aProg$ (resp. $\bProg$) over the common base~$\baseProg$ into a sequence of diff slots. 
At each slot, a change from either $\aProg$ or $\bProg$ is incorporated.
If both programs change a common slot, a {\it merge conflict} is produced, and requires manual resolution of the conflicting modifications.

\begin{figure}
% git merge-file --diff3 -p 1_a.js 1_base.js 1_b.js
\small{
\begin{tabular}{@{}llllll@{}}
%\begin{tabular}{@{}lllll@{}}
& \multicolumn{1}{c}{{Base $\baseProg$}} &
  \multicolumn{1}{c}{{Variant $\aProg$}} &
  \multicolumn{1}{c}{{Variant $\bProg$}} &
  \multicolumn{1}{c}{{Resolution?}} &\\
& \multicolumn{1}{c}{{\tt (base.js)}} & 
  \multicolumn{1}{c}{{\tt (a.js)}}  & 
  \multicolumn{1}{c}{{\tt (b.js)}} & 
  \multicolumn{1}{c}{{\tt (m.js)}} & \\
\hline
                 &  \\

\multirow{3}{*}{{\bf (1).}}
		 & {\tt\ y = 42;} & {\tt\ x = 1;}    & {\tt\ y = 42;} & {\tt\ x = 1;}\\
		 & 		  & {\tt\ y = 42;}   & {\tt\ z = 43;} & {\tt\ y = 42;}\\
		 & 		  &  		     & 		      & {\tt\ z = 43;}\\
                 &  \\
\hline
                 &  \\
\multirow{2}{*}{{\bf (2).}}
                 %& {\tt\ y = 42;}  & {\tt\ x = 1;}  & {\tt\ z = 43;} &  \multirow{2}{*}{{\tt Yes}} \\
		 & {\tt\ y = 42;}  & {\tt\ x = 1;}  & {\tt\ z = 43;} & {\tt\ CONFLICT}  \\
                 &                 & {\tt\ y = 42;} & {\tt\ y = 42;} &  \\
                 &  \\
\hline
\end{tabular}}
%\vspace{-0.1in}
\caption{Two examples of unstructured merges. \label{fig:intro:example}}
%\vspace{-0.2in}
\end{figure}

Figure~\ref{fig:intro:example} shows two simple code snippets to illustrate examples of three-way merge inputs and outputs. 
The figure shows the base program file $\baseProg$ along with the two variants $\aProg$ and $\bProg$. 
Example (1) shows a case where \diffThree{} successfully provides a merged file $\resProg$ incorporating changes made in both $\aProg$ and $\bProg$. 
On the other hand, Example (2) shows a case where \diffThree{} declares a conflict because two independent changes (updates to {\tt x} and {\tt z}) occur in the same diff slot. 

When \diffThree{} declares a conflict, a developer must intervene. 
Consequently, merge conflicts are consistently ranked as one of the most taxing issues in collaborative, open-source software development, "especially for seemingly less experienced developers"~\cite{Gousios-icse16}. 
Merge conflicts impact developer productivity, resulting in costly broken builds that stall the continuous integration (CI) pipelines for several hours to days. 
The fraction of merge conflicts as a percentage of merges range from 10\% --- 20\% for most collaborative projects. 
In several large projects, merge conflicts account for up to 50\% of merges (see~\cite{ghiotto-tse18} for details of prior studies).

% structured and semi-structured merge deal with structure, cannot deal with semantics,
Merge conflicts often arise due to the unstructured \diffThree{} algorithm that simply checks if two changes occur in the same diff slot. 
For instance, the changes in Example (2), although textually conflicting, do not interfere semantically. 
This insight has inspired research to incorporate program structure and semantics while performing a merge. 
{\it Structured merge} approaches~\cite{Apel11, lessenich2015, ASE2019} and their variants treat merge inputs as abstract syntax trees (ASTs), and use tree-structured merge algorithms. 
However, such approaches still yield a conflict on merges such as Example (2) above, as they do not model program semantics and cannot safely reorder statements that have side effects.\footnote{We ran {\tt jdime}~\cite{lessenich2015} in {\tt structured} mode on this example after translating the code snippet to Java.} 
To make matters worse, the gains from structured approaches hardly transfer to dynamic languages, namely JavaScript~\cite{ASE2019}, due to the absence of static types. 
Semantics-based approaches~\cite{Yang1992,Sousa18} can, in theory, employ program analysis and verifiers to detect and synthesize the resolutions. 
However, there are no semantics-based tools for synthesizing merges for any real-world programming language, reflecting the intractable nature of the problem.
Current automatic approaches fall short, suggesting that merge conflict resolution is a non-trivial problem.

% insight
%Our key insight stems from previous large scale studies of (Java) projects in GitHub that demonstrate that a significant fraction (87\%) of merge conflict resolutions are exclusively constructed from lines in the input \cite{ghiotto-tse18}. 

This paper takes a {\it fresh} {\it data-driven} approach to the problem of resolving unstructured merge conflicts. Inspired by the abundance of data in open-source projects, the paper demonstrates how to collect a dataset of merge conflicts and resolutions. 

This dataset drives the paper's key insight: a vast majority (80\%) of resolutions do not introduce new lines. Instead, they consist of (potentially rearranged) lines from the conflicting region.
This observation is confirmed by a prior independent large-scale study of Java projects from GitHub ~\cite{Gousios-icse16}, in which 87\% of resolutions are comprised exclusively from lines in the input. In other words, a typical resolution consists of re-arranging conflicting lines without writing any new code. Our observation naturally begs the question: \textit{Are there latent patterns of rearrangement? Can these patterns be learned?}

This paper investigates the potential for learning latent patterns of rearrangement. Effectively, this boils down to the question:
\begin{quote}
{\it Can we learn to synthesize merge conflict resolutions?}
\end{quote}
Specifically, the paper frames merging as a sequence-to-sequence task akin to machine translation.

%We note that it would be desirable to have a knob to control the set of resolutions that are suggested, to accommodate a spectrum of users with different tolerance for incorrect suggestions. 
%Such a data-driven merge algorithm can be paired to complement the lightweight, language-agnostic, unstructured merge algorithms such as \diffThree{}, by learning to mimic how developers resolve conflicting merges.
%Guided by a dataset of merge conflicts and resolutions, we formulate the merge algorithm as a machine learning problem. 

%Merge conflict resolution is a tedious task.
%It is typically tedious to resolve merge conflicts. 
%A vast majority (80\%) of merge conflict resolutions in our dataset exclusively take lines from the input sequence without adding any new code. 
%\shuvendu{Please qualify with stats}
%\TODO{Say that even though devs don't really write new code, it is a non-trivial problem to solve!! Compared to structured approaches we do much better}
%Rather than attempting to synthesize a merge, other approaches aim to verify if a merge is conflicting.
%Sousa et al.~\cite{Sousa18} provide a tool to only verify if a merge is conflicting for a small subset of Java

\begin{figure}
\centering
\begin{subfigure}[t!]{120pt}
{\footnotesize
\begin{tabular}{l}
\text{\text{\{ unchanged lines (prefix) \}}} \\
{\tt <{}<{}<{}<{}<{}<} \\
\text{\textsf{\{ lines edited by $\mathcal{A}$ \}}} \\
{\tt |||||||} \\
\text{\textsf{\{ affected lines of base $\mathcal{O}$ \}}} \\
{\tt =======} \\
\text{\textsf{\{ lines edited by $\mathcal{B}$ \}}} \\
{\tt >{}>{}>{}>{}>{}>} \\
\text{\textsf{\{ unchanged lines (suffix)} \}}
\end{tabular}
}
\caption{Format of a conflict.}
\end{subfigure}
\begin{subfigure}[t!]{100pt}
{\footnotesize
\begin{verbatim}
  <<<<<< a.js
  x = 1;
  |||||| base.js
  =======
  z = 43;
  >>>>>> b.js
  y = 42;
\end{verbatim}}
\vspace{0.20in}
\caption{Instance of a conflict.}
\end{subfigure}
%\begin{subfigure}[t!]{70pt}
%{\footnotesize
%\begin{verbatim}
%  <<<<<< a.js
%  x = 1;
%  w = 42;
%  |||||| base.js
%  y = 42;
%  ======
%  z = 43;
%  w = 42;
%  >>>>>> b.js
%\end{verbatim}}
%\vspace{0.03in}
%\caption{Conflict (II).}
%\end{subfigure}
%\vspace{-0.1in}
\caption{Conflict format and an instance reported by \diffThree{}
on Example (2) from Figure \ref{fig:intro:example}.
	\label{fig:intro:conflict}}
%\vspace{-0.1in}
\end{figure}

%\paragraph{{Data-Driven Merge.}}

% identify spurious conflicts
% construct resolutions
% attach a confidence metric (since we don't have an oracle)

%Our approach is inspired by our observation that most conflicts in our dataset are resolved by re-arranging lines in the input without writing any new code. Consequently, we believe that merge conflict resolution is an attainable task for sequence-to-sequence models\shuvendu{Why do we need to say about our beliefs}. 
%We have implemented these ideas into a data-driven, probabilistic, merge tool \deepMergeTool{}. 

To formulate program merging as a sequence-to-sequence problem, the paper considers the text of programs $\aProg$, $\bProg$, and $\baseProg$ as the input sequence, and the text of the resolved program $\resProg$ as the output sequence. 
However, this seemingly simple formulation does not come without challenges. 
Section \ref{sec:eval} demonstrates an out of the box sequence-to-sequence model trained on merge conflicts yields very low accuracy.
In order to effectively learn a merge algorithm, one must:
\begin{enumerate}
\item represent merge inputs in a concise yet sufficiently expressive sequence;
\item create a mechanism to output tokens at the line granularity; and %construct a concise yet sufficiently expressive output vocabulary
\item localize the merge conflicts and the resolutions in a given file.
%\item obtain an localized dataset of conflicts and resolutions for training \shuvendu{motivate that otherwise we will need to generate the entire file as a sequence}
\end{enumerate}

%These challenges concern representing the merge inputs, constraining the output vocabulary, and obtaining unambiguous ground truth to train the model. 

To represent the input in a concise yet expressive embedding, the paper shows how to construct an edit aware sequence to be consumed by \deepMergeTool{}. These edits are provided in the format of {\tt diff3} which is depicted in Figure \ref{fig:intro:conflict}(a) in the portion between
markers ``{\footnotesize \verb+<<<<<<<+}`` and ``{\footnotesize \verb+>>>>>>>+}``. The input embedding is extracted from parsing the conflicting markers and represents $\aProg$'s and $\bProg$'s edits over the common base $\baseProg$. 

To represent the output at the line granularity, \tool's design is a form of a pointer network \cite{NIPS2015_29921001}. As such, \tool{} constructs resolutions by copying input lines, rather than learning to generate them token by token.  Guided by our key insight that a large majority of resolutions are entirely comprised of lines from the input, such an output vocabulary is sufficiently expressive.

Lastly, the paper shows how to localize merge conflicts and the corresponding user resolutions in a given file. 
This is necessary as our approach exclusively aims to resolve locations in which \diffThree{} has declared a conflict. 
As such, our algorithm only needs to generate the conflict resolution and not the entire merged file. 
Thus, to extract ground truth, we must localize the resolution for a given conflict in a resolved file. 
Localizing such a resolution region unambiguously is a non-trivial task. 
The presence of extraneous changes unrelated to conflict resolution makes resolution localization challenging. 
The paper presents the first algorithm to localize the resolution region for a conflict. This ground truth is essential for training such a deep learning model. 

%Finally, training a supervised model like \deepMergeTool{} requires ground truth in the form of useful resolutions of unstructured merge conflicts.
%However, raw merge data of resolutions in a code corpus poses a host of challenges including unambiguously localizing the manual resolutions, the presence of extraneous changes unrelated to conflict resolution, and the presence of numerous {\it trivial} resolutions that discard one of the two branches; learning from such data would lead to a model providing useless resolutions.
%We describe our design tradeoffs for dealing with these problems at scale, including an algorithm to localize resolution regions unambiguously. 

The paper demonstrates an instance of \deepMergeTool{} trained to resolve unstructured merge conflicts in JavaScript programs.
Besides its popularity, JavaScript is notorious for its rich dynamic features, and lacks tooling support. Existing structured approaches struggle with JavaScript~\cite{ASE2019}, providing a strong motivation for a technique suitable for dynamic languages. The paper contributes a real-world dataset of 8,719 merge tuples that require non-trivial resolutions from nearly twenty thousand repositories in GitHub. Our evaluation shows that, on a held out test set, \deepMergeTool{} can predict correct resolutions for 37\% of non-trivial merges. \deepMergeTool{}'s accuracy is a 9x improvement over a recent semistructured approach ~\cite{ASE2019}, evaluated on the same dataset. Furthermore, on the subset of merges with upto 3 lines (comprising 24\% of the total dataset), \deepMergeTool{} can predict correct resolutions with 78\% accuracy. 
%\shuvendu{what about large conflicts?}

%\TODO{Our evaluation shows that \deepMergeTool{} can predict correct resolutions with high precision ($72$\%) and modest recall ($34$\%) on our dataset, with high recall ($78$\%) on merges comprising of up-to 3 lines that in turn comprise $24$\% of the dataset.}
%\deepMergeTool{} resolves $\textbf{37.8}$\%  of unstructured merge-conflicts on this dataset.
%Moreover, for merges limited to 3 lines in the input constituting $\textbf{24}$\% of our overall dataset, \deepMergeTool{} resolves $\textbf{78.4}$\%  of conflicts.
%Further, the probabilistic nature of the tool allows us improve the precision to $\textbf{72}$\% while retaining $\textbf{34}$\% of correct resolutions; in other words the tool produces only 1 incorrect resolution for every 4 suggestions.
%This shows that \deepMergeTool{} offers potential to be a useful tool to help mitigate the problem of spurious merge conflicts in modern version control systems. 
% contributions
\emph{Contributions.}
In summary, this paper:
\begin{enumerate}[leftmargin=*]
\item is the first to define merge conflict resolution as a machine learning problem and identify a set of
challenges for encoding it as a sequence-to-sequence supervised learning problem (\S~\ref{sec:formulation}).
\item presents a data-driven merge tool \deepMergeTool{} that uses
  edit-aware embedding to represent merge inputs and
  a variation of pointer networks to construct the resolved program (\S~\ref{sec:deepmerge}).
\item derives a real-world merge dataset%\footnote{https://tinyurl.com/deepmerge-data-E37A8EB7093358 Provided for reviewers only as we are pending legal approval. Will be externally released upon acceptance.} 
for supervised learning by
  proposing an algorithm for localizing resolution regions (\S~\ref{sec:dataset}).
\item performs an extensive evaluation of \deepMergeTool{} on merge conflicts in real-world JavaScript programs. And, demonstrates that it can correctly resolve a significant fraction of unstructured merge conflicts with high precision and 9x higher accuracy than a structured approach. 
\end{enumerate}

%\input{motivating-examples.tex}
%\input{background}
% !TeX root = ./main.tex
%\section{Merge conflict resolution as a Sequence2Sequence task}
\section{Data-Driven Merge}
\label{sec:formulation}

\newenvironment{myquote}%
  {\list{}{\leftmargin=0.28in\rightmargin=0.28in}\item[]}%
  {\endlist}

We formulate program merging as a sequence-to-sequence supervised learning problem and
discuss the challenges we must address in solving the resulting formulation.

\subsection{Problem Formulation}
%A program $P$ is a finite sequence of $N$ segments
%$(P_N)_{N\in\mathbb{N}}$ and we assume a function
%$\segments((P_N)) = \setc{P_i}{i \in 1..N}$
%that computes a set from the program sequence.
%We drop the subscript over sequences when unambiguous.

A merge consists of a 4-tuple of programs
$(\aProg, \bProg, \baseProg, \resProg)$ where $\aProg$ and $\bProg$
are both derived from a common $\baseProg$, and $\resProg$ is the
developer resolved program. 

A merge may consist of one or more regions.
We define a {\it merge tuple} (($A, B, O$), $R$) such that
$A$, $B$, $O$ are (sub) programs that correspond to regions in
$\aProg$, $\bProg$, and $\baseProg$, respectively, and
$R$ denotes the result of merging those regions.
%The result is of one of two kinds: $R$ or $\bot$.
%It is the resolved region $R$ in $\resProg$ if
%$\segments(R) \subseteq (\segments(A) \cup \segments(B))$.
%On the other hand, if a developer introduced new segments,
%then the result is a conflict, which we denote using $\bot$. 
%In Section 4.2, we describe in further detail how introducing new segemnts
%can indicate a non-spurious conflict. 
Although we refer to $(A, B, O, R)$ as a merge tuple, we assume that the tuples also implicitly contain the programs that they came from as additional contexts (namely $\aProg, \bProg, \baseProg, \resProg$). 

\begin{definition}[Data-driven Merge]
\label{defn:data-merge}
Given a dataset of $M$ merge tuples,
$$D = \{(A^i, B^i, O^i, R^i) \}_{i=1}^{M}$$ 
% our goal is to learn a merge algorithm $\algo$ that maximizes:
a data-driven merge algorithm $\algo$ is a function that maximizes:
$$\sum_{i=1}^{M} \algo(A^i, B^i, O^i) = R^i$$
treating Boolean outcomes of the equality comparison as integer constants 1 (for ${\tt true}$) and 0 (for ${\tt false}$). 
\end{definition}
In other words, $\algo$ aims to maximize the number of merges from $D$.
Rather than constraining merge to exactly satisfy {\it all} merge tuples in $D$, we relax the objective to maximization. A perfectly satisfying merge function may not exist in the presence of a real-world {\it noisy} dataset $D$. For instance, there may be $(A^i, B^i, O^i, R^i) \in D$ and $(A^j, B^j, O^j, R^j) \in D$ for $i \neq j$, $A^i = A^j$, $B^i = B^j$, $O^i = O^j$ but $R^i \neq R^j$.  In other words, two merge tuples consist of the same edits but different resolutions.

%Note that this definition by itself does not account for the rather hard
%problem of formalizing developer ``intent''.
%This paper argues we can condition the learning of $\algo$ on the
%developer's resolutions inherent in $D$ and thus capture the intent
%intrinsic in the developer's actions.

{\bf Example 1.}
Figure \ref{fig:problem:example}(a) shows a merge instance
that we will use as our running example throughout.
This instance is formulated in our setting as the merge tuple
$(A,B,O,R)$ depicted in Figure \ref{fig:problem:example}(b).
%There are many ways to segment the programs denoted by $A$,
%$B$, $O$, $R$ with varying tradeoffs.
%For instance, using line-based segmentation, we have three
%lines in $\segments(A)$, two lines in $\segments(O)$, and
%one line in $\segments(B)$.
$R$ contains only lines occurring in
the input. The two lines in $R$ correspond to
the first line of $B$ and the third line of $A$.
For this example, the $R$ also incorporates the intents from both $A$ and $B$ intuitively, assuming $\texttt{b}$ does not appear in the rest of the programs. 
%Thus, $R$ captures the intent of both $A$ and $B$.
%
%We discuss different tokenization mechanisms, their tradeoffs,
%and the choice of tokenization in $\deepMergeTool$ in Section \ref{sec:input}.
$\Box$

\newcommand{\rulesep}{\unskip\ \vrule\ }

\begin{figure}
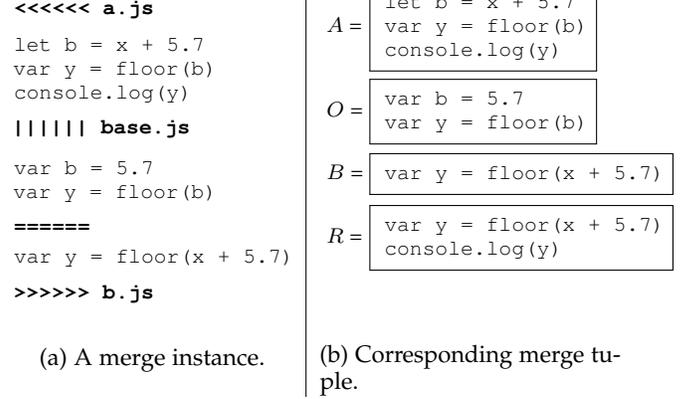

\centering
\begin{subfigure}[t!]{105pt}
{\footnotesize
\vspace{-0.05in}
{\bfseries\ttfamily <{}<{}<{}<{}<{}< a.js} \\
\vspace{-0.14in}
\begin{verbatim}
let b = x + 5.7
var y = floor(b)
console.log(y)
\end{verbatim}
\vspace{-0.02in}
{\bfseries\ttfamily |||||| base.js} \\
\vspace{-0.12in}
\begin{verbatim}
var b = 5.7
var y = floor(b)
\end{verbatim}
\vspace{-0.02in}
{\bfseries\ttfamily ======}
\vspace{-0.05in}
\begin{verbatim}
var y = floor(x + 5.7)
\end{verbatim}
\vspace{-0.05in}
{\bfseries\ttfamily >{}>{}>{}>{}>{}> b.js}
}
\vspace{0.15in}
\caption{A merge instance.}
\vspace{-0.12in}
\end{subfigure}
\
\rulesep
\ 
\begin{subfigure}[t!]{112pt}
\centering
{\footnotesize
\vspace{0.1in}
\begin{tabular}{@{\ }l@{\ }l@{}}
$A$ = & \fbox{
		\begin{tabular}{@{}l@{}}
			{\tt let b = x + 5.7}  \\
	 		{\tt var y = floor(b)} \\
			{\tt console.log(y)}
		\end{tabular}} \\[15pt]
$O$ = & \fbox{
		\begin{tabular}{@{}l@{}}
			{\tt var b = 5.7} \\
			{\tt var y = floor(b)}
		\end{tabular}} \\[11pt]
$B$ = & \fbox{
        \begin{tabular}{@{}l@{}}
		{\tt var y = floor(x + 5.7)}
		\end{tabular}} \\[7pt]
$R$ =  & \fbox{
        \begin{tabular}{@{}l@{}}
		{\tt var y = floor(x + 5.7)} \\
		{\tt console.log(y)} 
		\end{tabular}} 
\end{tabular}
}
\vspace{0.3in}
\caption{Corresponding merge tuple.}
\end{subfigure}
%\vspace{-0.1in}
\caption{Formulation of a merge instance in our setting.
\label{fig:problem:example}}
%\vspace{-0.1in}
\end{figure}

One possible way to learn a $\algo$ algorithm is by modeling the conditional
probability
\begin{equation}
  p(R | A, B, O)
  \label{eq:cond}
\end{equation}
In other words, a model that generates the output program $R$ 
given the three input programs.

Because programs are sequences, we further decompose Eq~\ref{eq:cond} by applying the chain rule~\cite{NIPS2014_a14ac55a}:
$$p(R | A, B, O) = \prod_{j=1}^N{p(R_j | R_{<j}, A, B, O)}$$ 
This models the probability of generating the $j-$th element of the program,
given the elements generated so far. 
There are many possible ways to model a three-way merge.  However, the
above formulation suggests one obvious approach is to use a maximum
likelihood estimate of a sequence-to-sequence model.

\subsection{Challenges}
\label{sec:challenges}
Applying a sequence-to-sequence (Seq2seq) model to merge conflict resolution
poses unique challenges.
We discuss three key challenges, concerning input representation, output construction,
and dataset extraction.

\subsubsection{Representing the Merge Inputs as a Sequence.}
In a traditional sequence-to-sequence task such as machine translation, there is a single input sequence that maps to a single output sequence. 
However, in our case, we have three input sequences of varying sizes, corresponding to the three versions of a program involved in a merge conflict.
It is not immediately evident how to determine a suitable token granularity
and encode these sequences in a manner that is amenable to learning.
One obvious solution is to concatenate the tokens of the three sequences
to obtain a single sequence. However, the order of concatenation is unclear. Furthermore, 
as we show in Section~\ref{sec:input},
such a naive representation not only suffers from information loss and
truncation, but also poor precision by being unaware of $A$ and $B$'s edits
over common base $O$.
%That section also develops an input tokenization scheme and an
%edit-aware input representation that are provably superior to alternatives
%in the literature and significantly increase our model efficacy.
In summary, we have:
\begin{myquote}
{\it \challenge{1}: Encode programs $A$, $B$, and $O$ as the input to a Seq2Seq~model.}
\end{myquote}

\subsubsection{Constructing the Output Resolution}
%\textbf{Output vocabulary} is too large - makes training and inference unrealisic
%Although a sequence-to-sequence model offers a natural framework for data-driven merge,
%it does not capture the highly restricted nature of the output sequence.
Our key insight that a majority of resolutions do not introduce new lines
leads us to construct the output resolution directly from lines in the conflicting region.
This naturally suggests the use of pointer networks~\cite{NIPS2015_29921001}, an encoder-decoder architecture capable of producing outputs explicitly pointing to tokens in the input sequence.
However, a pointer network formulation suggests an equivalent input and output granularity. 
In Section~\ref{sec:input}, we show that the input is best represented at a granularity far smaller than lines.

%This observation suggests restricting the vocabulary of the merged program $R$
%only to that of programs $A$ and $B$ as opposed to that of the
%underlying programming language (or even the entire input programs $\aProg$ and $\bProg$).

%This implies that a model that has an unrestricted output vocabulary
%over any possible token in the underlying programming language (or
%even restricted to the tokens in the input programs) will find it hard
%to capture the semantics of merge.
%
%Section~\ref{sec:output} discusses a way to restrict a model to just
%tokens from the input called $\segments$.
%However, it is important to note that even with such a restriction,
%the space of possible solutions is still large. 
% The space of possible solutions is large even for the restricted problem. 
% Assume the number of segments introduced in $A$ (resp. $B$)
% over the common base $O$ is $n$ (resp. $m$).
% Even when the merge is restricted to segments introduced by $A$ and
% $B$, there can still be $(n+m)^{n+m}$ possible output sequences.
% Intuitively, an output sequence is bounded by $n + m$ segments
% (ignoring repeated segments), and each segment can come from either $A$ or $B$.
% Even a simple scenario with 10 segments yields $10^{10} = 10$ billion sequences!
%\TODO{On the other hand, most such sequences would perhaps be syntactically ill-formed.[ONLY IF WE HAVE PARSER]}
%
Thus, the challenge is:
\begin{myquote}
  %{\it \challenge{2}: Encode merged program $R$ as the output of a \linebreak Seq2Seq model.}
  {\it \challenge{2}: Output $R$ at the line granularity given a non-line granularity input.} 
\end{myquote}

\subsubsection{Extracting Ground Truth from Raw Merge Data.}
Finally, to learn a data-driven merge algorithm, we need 
real-world data that serves as ground truth.
Creating this dataset poses non-trivial challenges.
First, we need to localize the resolution region and corresponding conflicting region.
In some cases, developers performing a manual merge resolution made changes unrelated to the merge. 
Localizing resolution regions unambiguously from input programs is challenging due to
the presence of these unrelated changes. 
Second, we need to be able to recognize and subsequently filter merge resolutions that do not incorporate both the changes.
%Second, we need to be able to recognize and subsequently filter those merge resolutions that contain unrelated refactorings done by the developer during merge. \shuvendu{is this related to the problem of localizing the resolution region?}
%\shuvendu{What about the problem of trivial merges?}
In summary, we have:
\begin{myquote}
  {\it \challenge{3}: Identify merge tuples $\{(A^i, B^i, O^i, R^i) \}_{i=1}^{M}$ given $(\aProg, \bProg, \baseProg, \resProg)$.}
\end{myquote}

%It is not obvious whether one considers the set of merges performed
%successfully by \diffThree, as they cause build or test failures later. 
%On the other hand, while resolving merge conflicts introduced by
%\diffThree, developers often make unrelated refactorings or bug-fixes
%that makes it hard to extract the resolution. 
%Section \ref{sec:dataset} describes our overall methodology and 
%resolution extraction algorithm to create a dataset in a manner
%that avoids common pitfalls.

% !TeX root = ./main.tex
\section{The \deepMergeTool{} Architecture}
\label{sec:deepmerge}

\def\STOP{\ensuremath{\langle {\tt STOP} \rangle}}

% model and arch overview
% - merge2matrix -> bidirectionalgru -> merge2lines(decoder + self-attention)

% merge2matrix : solving challenge 1
% - tokenization
% - alignment

% merge2lines : solving challenge 2
\begin{figure*}
\centering
%\vspace{-3mm}
\includegraphics[width=0.9\textwidth]{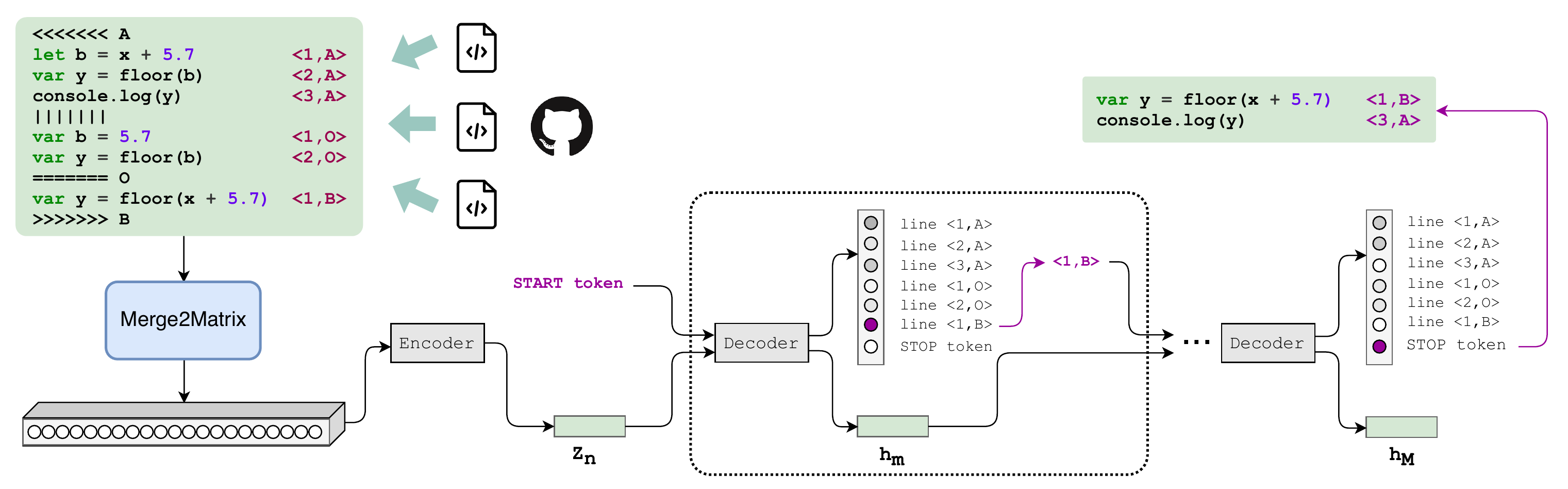}
%\vspace{-0.2in}
%https://app.diagrams.net/#G1BojFe__tzzVEbcRXbSaHKl5wNTcYW4h2
\caption{Overall \deepMergeTool{} framework. The dotted box represents repetition of \textit{decode} until $m = M$ i.e. the \STOP{} token is predicted. In this example, we have omitted $m=2$ in which the call to \textit{decode} outputs $y_2 = \langle 3,A \rangle.$ \label{fig:overall_framework}}
%\vspace{-0.05in}
\end{figure*}

Section~\ref{sec:formulation} suggested one way to learn a three-way
merge is through a maximum likelihood estimate of a
sequence-to-sequence model.  In this section we describe
\deepMergeTool{}, the first data-driven merge framework, and discuss
how it addresses challenges \challenge{1} and \challenge{2}.  We
motivate the design of \tool{} by comparing it to a standard
sequence-to-sequence model, the encoder-decoder architecture.

\subsection{Encoder Decoder Architectures\label{sec:enc_dec}}
%\TODO{this section is a bit confusing. Especially for someone who is not an expert in seq2seq models...}
Sequence-to-sequence models aim to map a fixed-length input ($(X_N)_{N\in\mathbb{N}}$), 
to a fixed-length output, ($(Y_M)_{M\in\mathbb{N}}$). \footnote{Note that $M$ is not necessary equal to $N$.}
The standard sequence-to-sequence model consists of three components: an \textbf{input embedding}, an \textbf{encoder}, and a \textbf{decoder}.

\textbf{Input embedding}: An embedding maps a discrete input from an input vocabulary $V$ ($x_n \in \mathbb{N}^{|V|}$), to a continuous $D$ dimensional vector space representation ($\overline{x}_n \in \mathbb{R}^D$). 
Such a mapping is obtained by multiplication over an embedding matrix $E \in \mathbb{R}^{D\times|V|}$. Applying this for each
element of $X_N$ gives $\overline{X}_N$. \\

%For each $x_n\in X_N$, an embedding
%extracts a column-vector of size $D$ from a matrix
%$E \in \mathbb{R}^{D\times|V|}$
%$$\overline{x}_n = E \cdot \textit{onehot}(x_n)$$ where
%$\mathit{onehot}$ maps a discrete token from an input vocabulary, $V$, 
%into its one-hot column-vector representation.  Applying this for each
%element of $X_N$ gives $(\overline{X}_N)_{N\in\mathbb{N}}$.
%

\textbf{Encoder}: An encoder \textit{encode}, processes each $\overline{x}_n$ and produces a hidden state, $z_n$ which summarizes the sequence upto the $n$-th element.
At each iteration, the encoder takes as input the current sequence element $x_n$, and the previous hidden state $z_{n-1}$. 
After processing the entire input sequence, the final hidden state, $z_N$, is passed to the decoder. \\ 

%Next, an encoder processes each $\overline{x}_n \in \overline{X}_N$
%and computes
%$$z_n = \textit{encode}(z_{n-1}, \overline{x}_n)$$ where $z_n$ is the
%hidden state of the encoder after the $n$-th element of the input sequence.
%
\textbf{Decoder}: A decoder \textit{decode}, produces the output sequence $Y_M$ from an encoder hidden state $Z_n$. 
Similar to encoders, decoders work in an iterative fashion. At each iteration, the decoder produces a single output token $y_m$
along with a hidden summarization state $h_m$. The current hidden state and the previous predicted token $y_{m}$ are then used in the following iteration to produce $y_{m+1}$ and $h_{m+1}$.
Each $y_m$ the model predicts is selected through a softmax over the hidden state:
$$p(y_m|y_1,...,y_{m-1},X) = \textit{softmax}(h_m)$$

%Given the final hidden state of the encoder ($z_N$), the decoder
%produces output tokens $y_m \in Y_M$
%\begin{align*}
%  h_1 &= z_N\\
%  h_m &= \textit{decode}(h_{m-1}, y_{m-1})
%\end{align*}
%where $h_m$ is the state of the decoder at the $m$-th element of the
%output sequence and $y_{m-1}$ is the prior predicted output token.
%

%
\deepMergeTool{} is based on this encoder-decoder architecture with
two significant differences.  

First, rather than a standard embedding followed by encoder, we
introduce a novel embedding method called \textit{Merge2Matrix}.
\textit{Merge2Matrix} addresses \challenge{1} by summarizing input programs ($A, B, O$) into a single embedding fed to the encoder. We discuss our \mergetomatrix solution as well as
less effective alternatives in Section~\ref{sec:input}.

Second, rather than using a standard decoder to generate output tokens
in some output token vocabulary, we augment the decoder to function as a variant of pointer networks.
The decoder outputs line tuples ($i$, $W$) where $W \in \{A, B\}$ and $i$ is the $i$-th line
in $W$. We discuss this in detail in
Section~\ref{sec:output}. 

{\bf Example 2.}  Figure~\ref{fig:overall_framework} illustrates the
flow of \tool{} as it processes the inputs of a merge tuple.  First, the raw text of $A, B,$ and $O$ is  fed to \mergetomatrix{}. As the name suggests, \mergetomatrix{} summarizes the tokenized inputs as a matrix. That matrix is then fed to an
encoder which computes the encoder hidden state $z_N$. Along with the
start token for the decoder hidden state, the decoder takes $z_N$ and
iteratively (denoted by the $\cdots$) generates as output the lines to
copy from $A$ and $B$.  The final resolution is shown in the green box.
$\Box$

\subsection{\mergetomatrix} \label{sec:input}
% \label{sec:merge2matrix}
% \begin{figure}
% \centering
% %\vspace{-3mm}
% \includegraphics[width=.4\textwidth]{figs/InputRepQuadrants}
% \vspace{-2mm}
% %https://app.diagrams.net/#G1wvoqi2nuhlDjh3tcIUiN123hXCseoln2
% \caption{Choices for Input Representation. \label{fig:input_rep_quad}}
% %\vspace{-4mm}
% \end{figure}
An encoder takes a single sequence as input. As discussed in
Section~\ref{sec:challenges}, a merge tuple consists of three sequences. This
section introduces \mergetomatrix, an input representation that
expresses the tuple as a single sequence. It consists of embedding,
transformations to summarize embeddings, and finally, edit-aware
alignment.

\subsubsection{Tokenization and Embedding}
This section discusses our relatively straightforward application of
both tokenization and embedding.

\emph{Tokenization.}
Working with textual data requires tokenization whereby we split a
sequence of text into smaller units referred to as
\textit{tokens}. Tokens can be defined at varying granularities such as 
characters, words, or sub-words. These units form a \textit{vocabulary} which maps input
tokens to integer indices. Thus, a vocabulary is a mapping from a
sequence of text to a sequence of integers.  This paper uses byte-pair
encoding (BPE) as it has been shown to work well with source code, where tokens can be formed by combining different words via casing conventions (e.g. $\texttt{snake\_case}$ or $\texttt{camelCase}$) causing a blowup in vocabulary size~\cite{10.1145/3377811.3380342}.  Byte-pair encoding is an unsupervised
sub-word tokenization that draws inspiration from information theory
and data compression wherein frequently occurring sub-word pairs are
recursively merged and stored in the vocabulary. We found that the performance of BPE was empirically superior to other tokenization schemes.

%We experimented with other tokenization schemes, but found that BPE was emperically best...

\emph{Embedding.}
Given an input sequence $X_N$, and a hyperparameter
(embedding dimension) $D$, an embedding transformation creates
$\overline{X}_{N}$. As described in Section~\ref{sec:enc_dec}, the
output of this embedding is then fed to an encoder.
% $P=(P_1,...,P_N)$ and converts each token in that input to a
% continuous representation $\overline{P}$. 
% Suppose $x_i$ is a one-hot vector which represents the $i$-th input token of $P$, then
% $$\overline{P}_i = E \cdot x_i$$ where $E \in \mathcal{R}^{D \times |V|}$,
% and $\overline{P} \in \mathcal{R}^{D \times N}$. The output of an embedding, $\overline{P}$ is then fed to the encoder.  
Because a merge tuple consists of three inputs ($A$, $B$, and $O$),
the following sections introduce novel transformations
that \emph{summarize} these three inputs into a format suitable for
the encoder.

\subsubsection{Merge Tuple Summarization}

In this section, we describe summarization techniques that are employed after embedding.
Before we delve into details, we first introduce
two functions used in summarization.

Suppose a function that concatenates embedded representations:
%$$ concat_3 : (\mathcal{R}^{D \times N} \rightarrow \mathcal{R}^{D
%\times N} \rightarrow \mathcal{R}^{D \times N}) \rightarrow
%\mathcal{R}^{D \times 3N}$$
$$ concat_s : (\mathbb{R}^{D \times N} \times \cdots \times \mathbb{R}^{D \times N}) \rightarrow \mathbb{R}^{D \times sN}$$
that takes $s$ similarly shaped tensors as arguments and concatenates
them along their last dimension. Concatenating these $s$ embeddings increases the size
of the encoder's input by a factor of $s$. 

%Long inputs to sequence-to-sequence models often suffer from information loss. Thus, 
%we posit it is better to \emph{learn} a linear composition of input sequences.  
Suppose a function
$linearize$ that linearly combines $s$ embedded representations. We
parameterize this function with learnable parameters
$\theta \in \mathbb{R}^{s+1}$. As input, $linearize$ takes an
embedding $\overline{x}_i \in \mathbb{R}^{D}$ for $i \in 1..S$.  Thus,
we define
$$linearize_{\theta}(\overline{x}_1, \dots, \overline{x}_s) = \theta_1 \cdot \overline{x}_1 + \cdots + \theta_s \cdot \overline{x}_s + \theta_{s+1}$$
where all operations on the inputs $\overline{x}_1,\dots,\overline{x}_s$ are
pointwise. $linearize$ reduces the size of the embeddings fed to
the encoder by a factor of $s$.

Now that we have defined two helper functions, we describe two summarization methods. %It is assumed that each summarization method is invoked \textit{after}
%tokenizing and embedding $A$, $B$, and $O$ to produce  $\overline{A}$, $\overline{B}$, and $\overline{O}$

\textbf{\emph{Na\"ive.}}
Given a merge tuple's inputs
($A, B, O$), a \naive{} implementation of \mergetomatrix is to 
simply concatenate the embedded representations
(i.e., $concat_3(\overline{A}, \overline{B}, \overline{O})$) 
% discuss issues with long input and figure quad
%
Traditional sequence-to-sequence models often suffer from information
forgetting; as the input grows longer, it becomes harder for $\textit{encode}$ to capture long-range correlations in that input.  % Thus, when training we limit
A solution that addresses \challenge{1}, must be concise  
while retaining the information in the input programs.  
% Denote the
% lengths of the inputs A, B, O after tokenization and embedding using $L_A, L_B, L_O$
% respectively. In our configuration, we set $N_{ctx} = 1024$. Each of
% our configurations in Figure~\ref{fig:input_rep_quad} handles inputs with representations longer than the $N_{ctx}$ through
% truncation, denoted in Figure~\ref{fig:input_rep_quad} by
% $\approx$. After $concat$ is applied, truncation occurs if $L_A + L_B + L_O > 1024$. 
% In these cases, the model fails to see the entirety of the
% input program and is forced to make a decision based on partial
% information. \shuvendu{lets downplay truncation based on the data and highlight that order of concat matters for different examples}

\textbf{Linearized.}
As an attempt at a more concise representation, we introduce a summarization we call
\emph{linearized}. This method linearly combines each of the embeddings through our helper function:
$linearize_{\theta}(\overline{A}, \overline{B}, \overline{O})$.
In Section~\ref{sec:eval} we empirically
demonstrate better model accuracy when we summarize with
$linearize_\theta$ rather than $concat_s$.

\subsubsection{Edit-Aware Alignment}
\begin{figure*}
\centering
\includegraphics[width=\textwidth]{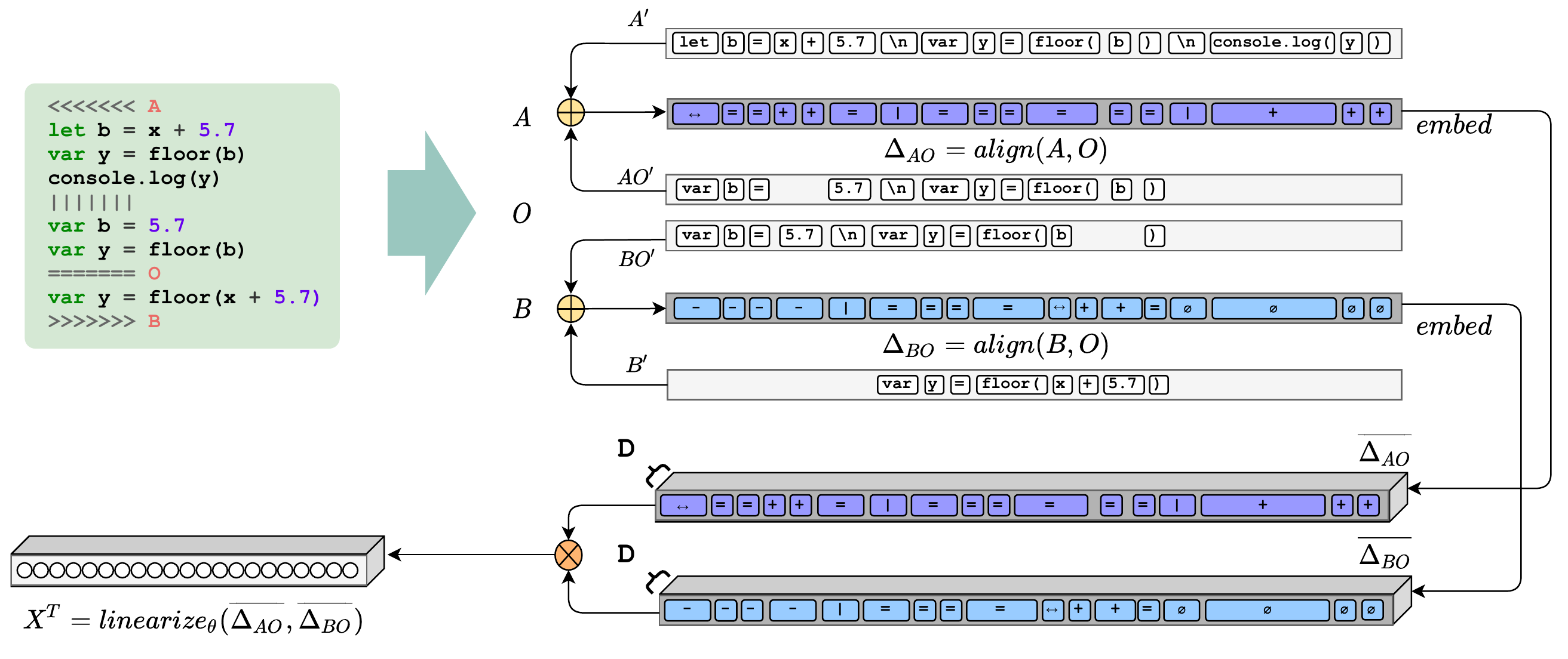}
%\vspace{-0.2in}
\caption{\textit{Merge2Matrix}: implemented with the Aligned Linearized input representation used in \deepMergeTool{}. \label{fig:InputRep}}
%\vspace{-0.1in}
\end{figure*}

In addition to input length, \challenge{1} also alludes that an
effective input representation needs to be ``edit aware''. The
aforementioned representations do not provide any indication that
$A$ and $B$ are edits from $O$.

Prior work, {\it Learning to Represent Edits} (LTRE)~\cite{yin2018learning} 
introduces a representation to succinctly encode 2 two-way diffs.
The method uses a standard deterministic diffing algorithm and represents the
resulting pair-wise alignment as an auto-encoded fixed dimension vector.

%The two-way alignment of  $A$ (resp. $B$) relative to $O$, computed by the function $align$, results in
%two token sequences of the same length, $A'$ and $AO'$ (resp $B'$ and $BO'$) , having unchanged tokens in the
%same positions and padding symbols otherwise.  
%In addition, a two-way
A two-way alignment produces an ``edit sequence''. This series of edits, 
if applied to the second sequence, would produce the first.
An edit sequence, $\Delta_{AO}$, is comprised of
the following editing actions: $\bold{=}$ representing equivalent tokens, $\bold{+}$
representing insertions, $\bold{-}$ representing deletions,
$\bold{\leftrightarrow}$ representing a replacement. Two special tokens
$\bold{\emptyset}$ and $\bold{\vert}$ are used as a padding token and a newline marker, respectively. %Define $\Delta_{AO}$
%(resp $\Delta_{BO}$) as the edit sequence of these actions from $O$ to $A$
%(resp $B$) and the function $align$ that computes them.
% %Next, we extract an edit
% %sequence corresponding to each pair of aligned sequences.  Following
% %the \ltre{} representation\cite{yin2018learning},
% The alignment of $A$ (resp. $B$) relative to $O$ results in two token
% sequences of the same length, having unchanged tokens in the same
% positions, and padding symbols otherwise. To compute that alignment, we introduce the
% following symbols to denote sequence editing actions: $\bold{=}$
% represents equivalent tokens, $\bold{+}$ represents insertions,
% $\bold{-}$ represents deletions, $\bold{\leftrightarrow}$ represents a
% replacement, and $\bold{\emptyset}$ is used as a padding
% token. Finally, because \tool{} copies lines from the input, we also
% augment \ltre's set of alignment tokens with $\bold{|}$ to denote when
% newlines occur on that input. % In this
% % way we can generate a single sequence representing $O$ and the edit
% % from $O$.
Note that these $\Delta$s only capture information about the {\it
  kinds} of edits and \emph{ignore} the the tokens that make up the
edit itself (with the exception of the newline token).
Prior to the creation of $\Delta$, a preprocessing step adds padding tokens
such that equivalent tokens in A (resp. B) and O
are in the same position. These sequences, shown in 
Figure~\ref{fig:InputRep} are denoted as $A'$ and $AO'$ (resp. $B'$ and 
$BO'$).

{\bf Example 3.}  Consider $B$'s edit to $O$ in
Figure~\ref{fig:InputRep} via its preprocessed sequences $B'$, $BO'$,
and its edit sequence $\Delta_{BO}$.
One intuitive view of $\Delta_{BO}$ is that it is a set of
instructions that describe how to turn $B'$ into $BO'$ with the
aforementioned semantics.  Note the padding token $\emptyset$
introduced into $\Delta_{BO}$ represents padding out to the length of
the longer edit sequence $\Delta_{AO}$.
% where $\bold{+}$
% denotes addition, $\bold{=}$ denotes a match, $\bold{-}$ denotes
% deletion, $\bold{\leftrightarrow}$ denotes a replacement, and
% $\bold{|}$ denotes a newline. 
$\Box$

We now describe two edit-aware summarization methods based on this
edit-aware representation. However, our setting differs from the 
original \ltre{} setting as we assume three input sequences 
and a three-way diff. In the following summarization methods, 
we assume that $A, B, O$ are tokenized, but \textit{not} embedded 
before invoking \mergetomatrix{}.

% Let $L_{AO}$ and $L_{BO}$ be the length of $\Delta_{AO}$ and
% $\Delta_{BO}$, respectively, where $\Delta_{AO}$ and $\Delta_{BO}$ are
% the diff sequences constructed from $A$ and $O$, and $B$ and $O$,
% respectively. \elizabeth{Add something about newline markers}

\textbf{Aligned \emph{na\"ive}.}
Given $\Delta_{AO}$ and $\Delta_{BO}$, we embed each to produce
$\overline{\Delta_{AO}}$ and $\overline{\Delta_{BO}}$, respectively.
Then we combine these embeddings through concatenation and thus
$concat_2(\overline{\Delta_{AO}}, \overline{\Delta_{BO}})$ is fed to
the encoder.  % As in the edit-unaware concatenation described earlier, when
% $L_{AO} + L_{BO} > N_{ctx}$, truncation occurs and the model is again forced
% to make a decision based on a partial information. \shuvendu{downplay truncation}

\textbf{Aligned linearized.} 
This summarization method is depicted in Figure~\ref{fig:InputRep}, 
invoking $linearize$ to construct an input representation over
edit sequences.  
First, we apply alignment to create $\Delta_{AO}$ and
$\Delta_{BO}$. This is portrayed through the $\bigoplus$
operator. Following construction of the $\Delta$s, we apply embedding
and subsequently apply our edit-aware linearize operation via the
$\bigotimes$ operator. Thus, we summarize embeddings with
$linearize_{\theta}(\overline{\Delta_{AO}},\overline{\Delta_{BO}})$
and feed its output to the encoder.  As we demonstrate in
Section~\ref{sec:eval}, this edit-aware input representation
significantly increases the model's accuracy.

\textbf{\emph{LTRE.}} Finally, for completeness, we also
include the original \ltre{} representation. We modify this to our
setting by creating \textit{two} 2-way diffs. The original \ltre{} 
has a second key difference from our summarization methods. 
\ltre{} includes all tokens from from the input sequences in addition to 
the edit sequences That is, \ltre{} summarizes $A'$
$AO'$, $\Delta_{AO}$, $B'$, $BO'$, and $\Delta_{BO}$. Let $\overline{A'}$, $\overline{AO'}$ and
$\overline{\Delta_{AO}}$, (resp $\overline{B'}$, $\overline{BO'}$, and
$\overline{\Delta_{BO}}$) be the embedding of a two-way diff.  Then,
the following summarization combines all embeddings:
$$concat_6(\overline{\Delta_{AO}}, \overline{A'}, \overline{AO'}, 
\overline{\Delta_{BO}}, \overline{B'}, \overline{BO'})$$

\subsection{The Encoder}\label{sec:encoder}
The prior sections described \mergetomatrix which embeds a merge into
a continuous space which is then summarized by an encoder.  % The
% encoder takes as input an embedded sequence
% $\overline{P} = (\overline{P}_1, ..., \overline{P}_N)$ and produces
% hidden vectors $Z = (Z_1,...,Z_N)$.
\tool{} uses a bi-directional
gated recurrent unit\cite{cho2014learning} (GRU) to summarize the embedded input
sequence.  We empirically found that a bi-directional GRU was more
effective than a uni-directional GRU.

\subsection{Synthesizing Merge Resolutions }\label{sec:output}
This section summarizes \tool's approach to solving \challenge{2}.
% and takes advantage of the empirical observation that 87\% of developer
%resolutions copy lines from the edits of $A$ or
%$B$~\cite{ghiotto-tse18}.
% In the standard NLP task of machine translation (for instance, English
% to German translation), given a $N$ token sequence in the source
% language (input vocabulary) the training objective is to generate
% $M^{'}$ tokens in the target language (output vocabulary). In this
% case, most tokens are generated in the target language, but rare
% elements, such as names, are copied from the input sequence. In the
% case of three-way program merge, most of the lines of code remain
% unchaged, and need to be reproduced in the output sequence. In this
% case, decoding sequences token-by-token is too hard and not needed.
Given a sequence of hidden vectors $Z_N$ produced by an encoder, a
decoder generates output sequence $Y_M$.
%Because merge resolutions are often lines copied directly from the input, 
We introduce an extension of a traditional decoder to
copy lines of code from those input programs.

Denote the number of lines in $A$ and $B$ as $Li_{A}$ and $Li_{B}$, respectively.
Suppose that $L = 1..(Li_{A} + Li_{B})$; then, a value $i \in L$
corresponds to the $i$-th line from $A$ if $i <= Li_{A}$, and the $i - Li_{A}$-th line
from $B$, otherwise.

Given merge inputs ($A$, $B$, $O$), \tool's decoder computes
a sequence of hidden states $H_M$, and models the conditional
probability of \emph{lines} copied from the input programs $A$, $B$,
and $O$ by predicting a value in $y_m \in Y_M$:
$$p(y_m|y_1,...,y_{m-1},A,B,O) = \textit{softmax}(h_m)$$ where $h_m
\in H_M$ is the decoder hidden state at the $m$-th element of the
output sequence and the $\textit{argmax}(y_m)$ yields an index into $L$.

% Note that our model predicts merge resolutions up to $C$ lines. We set $C=30$
% to tackle implementation constraints and because most resolutions are
% less than 30 lines long. However, we evaluate \deepMergeTool{} on 
% a full test dataset including samples with $M \geq C$.
% Assuming that the encoder defined in Section~\ref{sec:encoder} produces hidden
% states $Z = (Z_1,...,Z_N)$ we implement this conditional probability
% as:
% %\begin{equation}
% %K = \texttt{softmax}([\begin{array}{cc}\overline{l_{i-1}} & Z_{N}\end{array}] \cdot \theta)
% %\end{equation}
% \begin{equation}
% p(l_{i} | l_{<i}, A, B, O) = \prod_{i=1}^{C}{\mathbb{D}(l_{i-1}, Z)}
% %\texttt{argmax}(\texttt{softmax}(ll(K \cdot Z)))
% %p(l_{i} | l_{<i}, A, B, O) = \texttt{softmax}(\mathbb{D}(Z \cdot K)
% \end{equation}
% %$Z_N: (1 \times D)$ \\
% %$\overline{l_{i-1}}: (1 \times D)$  \\
% %$[\begin{array}{cc}\overline{l_{i-1}} & Z_{N}\end{array}]: (1 \times 2*D)$ \\
% %$\theta: (2*D \times N)$ \\
% %where $K \in \mathbb{R}^{1 \times N}$ and $ll: \mathbb{R}^{1 \times N} \rightarrow \mathbb{R}^{|M|+2}$
% %is the linear attention layer, 
% where the output of $\mathbb{D}(l_{i-1}, Z) \in \mathbb{R}^{|M|+2}$,
% and $|M|$ is the number of lines in the resolution. 
% \shuvendu{Don't understand $\mathbb{D}$}
In practice, we add an additional \STOP{} token to $L$. 
The \STOP{} token signifies that the decoder has completed the sequence.  
The \STOP{} token is necessary as the decoder may
output a variable number of lines conditioned on the inputs.

This formulation is inspired by pointer
networks~\cite{NIPS2015_29921001}, an encoder-decoder
architecture that outputs an index that explicitly points to an input
token.  Such networks are designed to solve combinatorial problems like
sorting. Because the size of the output varies as a function of the
input, a pointer network requires a novel attention mechanism that applies
attention weights directly to the input sequence. This differs from 
traditional attention networks which are applied to the outputs of the 
encoder $Z_N$. In contrast, \tool{} requires no change to attention. 
Our architecture outputs an index that points to the abstract 
concept of a line, rather than an explicit token in the input.
Thus, attention applied to $Z_N$, a summarization of the input,
is sufficient.

\subsection{Training and Inference with \deepMergeTool}

The prior sections discussed the overall model architecture of
\deepMergeTool. This section describes hyperparameters that control
model size and how we trained the model.
We use a embedding dimension $D=1024$ and 1024 hidden units in the
single layer GRU encoder.  %In total, \tool{} has 131,722,272 parameters.
Assume the model
parameters are contained in $\theta$; training seeks to find the
values of $\theta$ that maximize the log-likelihood
$$\argmax_{\theta} \log p_{\theta}(R | A, B, O)$$
over all merge tuples (($A, B, O$), $R$) in its training
dataset.  We use standard cross-entropy loss with the Adam optimizer.
Training takes roughly 18 hours on a NVIDIA P100 GPU and we pick the
model with the highest validation accuracy, which occurred after 29
epochs.  

Finally, during inference time, we augment \tool{} to use standard
beam search methods during decoding to produce the most likely $k$ top
merge resolutions. \tool{} predicts merge resolutions up
to $C$ lines. We set $C=30$ to tackle implementation constraints and
because most resolutions are less than 30 lines long. However, we
evaluate \deepMergeTool{} on a full test dataset including samples
where the number of lines in $M$ is $\geq C$. 

\section{Real-World Labeled Dataset}
\label{sec:dataset}

\newcommand{\words}[1]{\it tokens(#1)}

This section describes our solution to \challenge{3}:
localizing merge instances $(A, B, O, R)_i$ from $(\aProg, \bProg, \baseProg, \resProg)$.
Since a program may have several merge conflicts, we decompose the overall merge problem into merging individual instances. 
As shown in Figure~\ref{fig:problem:example}, $A$, $B$, and $O$ regions can be easily extracted given the \diffThree{} conflict markers.
However, reliably localizing a resolution $R$ involves two sub-challenges: 
\begin{enumerate}[leftmargin=*]
\item How do we localize individual regions $R$ unambiguously?
\item How do we deal with trivial resolutions?
\end{enumerate}
In this section, we elaborate on each of these sub-challenges and discuss our solutions. We conclude with a discussion of our final dataset and its characteristics. \\

\newcommand{\mt}{\it MT}
\newcommand{\nil}{{\tt nil}}
\newcommand{\stL}[1]{{\it spos}}
\newcommand{\enL}[1]{{\it epos}}
\newcommand{\prefixStr}{\it prfx}
\newcommand{\suffixStr}{\it sffx}
\newcommand{\bof}{{\it \ensuremath{\langle} BOF \ensuremath{\rangle}}}
\newcommand{\eof}{{\it \ensuremath{\langle} EOF \ensuremath{\rangle}}}
\newcommand{\Length}{\it Length}
\newcommand{\lines}{\textsc {Lines}}

\renewcommand{\algorithmicrequire}{\textbf{Input:}}
\renewcommand{\algorithmicensure}{\textbf{Output:}}

\begin{algorithm}
{\footnotesize
\caption{Localizing Merge Tuples from Files for Dataset}
\label{algo:dataset}
\begin{algorithmic}[1]
\Procedure{LocalizeMergeTuples}{$\mergeProg$, $\resProg$}
\State $\mt \gets \emptyset$ \Comment{Merge Tuples}
\For {$i \in $ [1, \textsc{NumConflicts}$(\mergeProg)$]}
    \State $R \gets $ \textsc{LocalizeResRegion}$(\mergeProg, \resProg, i)$
    \If {$R == \nil$}  \label{line:alg1:fail}
       \State {\bf continue}   \Comment{Could not find resolution}
    \EndIf
    \State $(A, B, O) \gets $\textsc{GetConflictComponents}$(\mergeProg, i)$
    \If {$R \in \{A, B, O\}$} \label{line:alg1:trivial}
         \State \textbf{continue} \Comment{Filter trivial resolutions}
    \EndIf \label{line:alg1:endfilter}
    %\EndIf \label{line:alg1:endfilter}
    \If {$\lines(R) \subseteq \lines(A) \cup \lines(B)$} 
    	\State $\mt \gets \mt \cup \{(A, B, O, R)\}$ \label{line:alg1:positive}
    %\ElsIf {$|\words{R}\setminus\words{A}\setminus\words{B}| == 1$} \label{line:alg1:conflict}
    %    \State $\mt \gets \mt \cup \{(A, B, O, \bot)\}$ \Comment{Non-spurious}
    \EndIf \label{line:alg1:endfilter}
\EndFor  
\State \Return $\mt$
\EndProcedure

\

\Procedure{LocalizeResRegion}{$\mergeProg$, $\resProg$, $i$}
\State $n \gets \Length(\resProg)$ \Comment{Length of $\resProg$ in chars}
\State $m \gets \Length(\mergeProg)$ \Comment{Length of $\mergeProg$ in chars}
\State $(\stL{i}, \enL{i}) \gets$ \textsc{GetConflictStartEnd}($\mergeProg, i$)  \label{line:alg1:markers}
\State $\prefixStr \gets \ \bof \ + \ \mergeProg[0:\stL{i}]$ \label{line:alg1:prefix}
\State $\suffixStr \gets \mergeProg[\enL{i}:m]+ \eof$ \label{line:alg1:suffix}
\State s $\gets$ \textsc{MinimalUniquePrefix}(${\it reverse}(\prefixStr), {\it reverse}(\resProg)$)\label{line:alg1:startRes}
\State e $\gets$ \textsc{MinimalUniquePrefix}($\suffixStr, \resProg$)\label{line:alg1:endRes}
\If {$s \geq 0$ and $e \geq 0$} 
  \State \Return $\resProg[n - s:e]$ \label{line:alg1:region}
\Else
  \State \Return $\nil$
\EndIf
\EndProcedure

\ 

\Procedure{MinimalUniquePrefix}{$x$, $y$}
\State
\algorithmicensure{Returns the start position of the minimal non-empty prefix of $x$ that appears uniquely in $y$, else -1}
\EndProcedure

\

\Procedure{lines}{$p$}
\State
\algorithmicensure{Returns the set of lines comprising the program $p$}
\EndProcedure

\end{algorithmic}
}
\end{algorithm}
%\vspace{-5mm}

Algorithm~\ref{algo:dataset} denotes a method to localize merge tuples from a corpus of merge conflict and resolution files. The top-level procedure $\textsc{ExtractMergeTuples}$ takes $\mergeProg$, the \diffThree{} conflict file with markers, along with $\resProg$, the resolved file. From those inputs, it extracts merge tuples into $\mt$. The algorithm loops over each of the conflicted regions in $\mergeProg$, and identifies the input $(A, B, O)$ and output ($R$) of the tuple using $\textsc{GetConflictComponents}$ and $\textsc{LocalizeResRegion}$ respectively.
Finally, it applies a filter on the extracted tuple (lines~\ref{line:alg1:fail}  -- \ref{line:alg1:endfilter}). 
We explain each of these components in the next few subsections. 

\subsection{Localization of Resolution Regions}
\label{sec:extraction}

\begin{figure}
\centering
\begin{subfigure}[t!]{100pt}
{\footnotesize
\begin{verbatim}
<BOF>
...
var time = new Date();
print_time(time);
<<<<<<< a.js
x = foo();
||||||| base.js
=======
x = bar();
>>>>>>> b.js
print_time(time);
<EOF>
\end{verbatim}}
\caption{A merge instance.}
%\vspace{-0.03in}
\end{subfigure}
\ \ \ \ 
\rulesep
\ \ \ \
\begin{subfigure}[t!]{90pt}
%\vspace{0.35in}
{\footnotesize
\begin{verbatim}
<BOF>
...
let time = new Date();
print_time(time);
baz();
print_time(time);
<EOF>
\end{verbatim}}
%\vspace{0.22in}
\caption{Resolution.}
%\vspace{-0.03in}
\end{subfigure}
\caption{Challenging example for localizing resolution.
\label{fig:challenge-resolution}}
%\vspace{-0.2in}
\end{figure}

Creating a real-world merge conflict labeled dataset requires identifying the ``exact'' code region that constitutes a resolution. 
However, doing so can be challenging; Figure~\ref{fig:challenge-resolution} demonstrates an example.
The developer chooses to perform a resolution \texttt{baz();} that does not correspond to anything from the $A$ or $B$ edits, and the surrounding context also undergoes changes (e.g. changing \textrm{var} with \textrm{let} which restricts the scope in the prefix).
To the best of our knowledge, there is no known algorithm to localize $R$ for such cases. 

$\textsc{LocalizeResRegion}$ is our method that tries to localize the $i^{th}$ resolution region $R$, or returns $\nil$ when unsuccessful. 
Intuitively, we find a prefix and suffix in a merge instance and use this prefix and suffix to bookend a resolution.  
If we cannot uniquely find those bookends, we say the resolution is \emph{ambiguous}.  

The method first obtains the prefix $\prefixStr$ (resp. suffix $\suffixStr$) of the $i^{th}$ conflict region in $\mergeProg$ in line~\ref{line:alg1:prefix} (resp. line~\ref{line:alg1:suffix}).
We add the start of file $\bof$ and end of file $\eof$ tokens to the prefix and suffix respectively. 
%Note that the prefix and the suffix may contain other conflict markers but are assumed not to appear in the resolved file $\resProg$.
The next few lines try to match the prefix $\prefixStr$ (resp. suffix $\suffixStr$) in the resolved file $\resProg$ unambiguously. 
Let us first focus on finding the suffix of the resolution region in $\resProg$ in line~\ref{line:alg1:endRes}.
The procedure $\textsc{MinimalUniquePrefix}$ takes two strings $x$ and $y$ and finds the start position of the minimal non-empty prefix of $x$ that appears uniquely in $y$, or returns -1. 
For example, $\textsc{MinimalUniquePrefix}$(``abc'', ``acdabacc'') is 3 since ``ab'' is the minimal prefix of $x$ that appears uniquely in $y$ starting in position 3 (0-based indexing).

To find the prefix of the resolution, we reverse the $\prefixStr$ string and search for matches in reversed $\resProg$, and then finally find the offset from the start of $\resProg$ by subtracting $s$ from the length $n$ of $\resProg$.
The unique occurrence of both the prefix and suffix in $\resProg$ allows us to map the conflicted region to the resolved region. 

For our example, even though the line ``$\texttt{print\_time(time);}$'' that encloses the conflicted region appears twice in $\resProg$, extending it by ``\texttt{time = new Date();}'' in the prefix and $\eof$ in the suffix provides a unique match in $\resProg$. 
Thus, the algorithm successfully localizes the desired region ``\texttt{baz();}'' as the resolution region. 

After localizing the resolution regions, we have a set of merge instances of the form $(A, B, O, R)$. 
We can use our definition from Section~\ref{sec:formulation} to label a merge tuple $(A, B, O, R)$.

\subsection{Filtering Trivial Resolutions}
\label{sec:trivial}
Upon examining our dataset, we found a large set of merges in which $A$ was taken as the resolution and $B$ was entirely ignored (or vice versa). These trivial samples, in large, were the product of running {\it git merge}  with ``ours'' or ``theirs'' command-line options. Using these merge options indicates that the developer did not resolve the conflict after careful consideration of both branches, but instead relied on the git interface to completely drop one set of changes. The aforementioned command-line merge options are typically used the commit is the first of many fix-up commits to perform the full resolution. 

We appeal to the notion of a ``valid merge'' that tries to incorporate both the syntactic and semantic changes from both $A$ and $B$. Thus, these samples are not valid as they disregard the changes from $B$ (resp. $A$) entirely. Furthermore, these trivial samples comprised 70\% of our ``pre-filtering'' dataset. Previous work confirmed our observation that a majority of merge resolutions in GitHub Java projects (75\% in Table~13~\cite{ghiotto-tse18}) correspond to taking just $A$ or $B$. To avoid polluting our dataset, we filter such merges $(A, B, O, R)$ where $R \in \{A, B, O\}$ (line~\ref{line:alg1:trivial} in Algorithm~\ref{algo:dataset}). Our motivation to filter the dataset of trivial labels is based on both dataset bias and the notion of a valid merge. 

\subsection{Final Dataset}
We crawled repositories in GitHub containing primarily JavaScript files, looking at merge commits.
%Gathering data from GitHub indiscriminately can lead to both noise and bias.  
%We address this by following the advice of Kalliamvakou~\emph{et al.}~\cite{kalliamvakou2014promises}, 
To avoid noise and bias, we select projects that were active in the past one year (at the time of writing), and received at least 100 stars (positive sentiment). 
We also verified that the dataset did not contain duplicate merges.
%We chose JavaScript as the language of focus for evaluation due to its importance and growing popularity and the fact that static analysis of JavaScript is challenging due to its weak, dynamic type system and permissive nature~\cite{jensen2009type,kashyap2014jsai}. 
We ignore {\it minified} JavaScript files that compress an entire JavaScript file to a few long lines.
%Some of them have a ``.min.js'' filename extension, but we found several that just has a ``.js'' extension. 
%We therefore adopted the strategy to remove any file that contains more than 1000 chars in a single line. 
Finally, note that Algorithm~\ref{algo:dataset} filters away any resolution that consists of new segments (lines) outside of $A$ and $B$ as our technique targets resolutions that do not involve writing any new code. 
%Starting with 56,818 JavaScript files with conflicts from 18,515 repositories,  
After applying filters, we obtained 8,719 merge tuples.
%\shuvendu{Ensure these numbers are consistent in abstract/intro @Elizabeth} 
We divided these into a 80/10/10 percent training/validation/test split.  
%\TODO{describe the training/validation/test split}
Our dataset contains the following distribution in terms of total number of lines in $A$ and $B$: 45.08\% ([0,5]), 20.57\% ([6,10]), 26.42\% ([11,50]), 4.22\% ([51,100]) and 3.70\% (100+). 

%4,933 (43.88\%) samples less than or equal to 5 lines, 2,193 (19.51\%) less than or equal to 10 lines, but greater than 5, 2,774 (24.68\%) less than or equal too 50 lines, but greater than 10, 541 (4.81\%) samples less than or equal to 100 lines but greater than 50, and 800 (7.12\%) samples with greater than 100 lines. 

% !TEX root = ./main.tex
\section{Evaluation}
\label{sec:eval}

\def\myline#1{\ensuremath{\langle #1 \rangle}}
\def\scanMerge{{\sc ScanMerge}}
\def\fairseq{{\sc Seq2Seq}}
\def\fstMerge{{\sc jsFSTMerge}}
\def\segments{\mathit{segments}}
\def\concat{\texttt{CONCAT}}
\def\other{\texttt{OTHER}}

In this section, we empirically evaluate \deepMergeTool{} to answer the following questions:
%\footnote{We will make data available upon acceptance.}
%We are working on the data release approval, and 

\begin{enumerate}
\item[\textbf{RQ1}] How effective is \deepMergeTool{} at synthesizing resolutions?
\item[\textbf{RQ2}] How effective is \deepMergeTool{} at suppressing incorrect resolutions? 
\item[\textbf{RQ3}] On which samples is \deepMergeTool{} most effective?
\item[\textbf{RQ4}] How do different choices of input representation impact the performance of \deepMergeTool{}?
\end{enumerate}

\subsection{RQ1: Effectiveness of Resolution Synthesis}
%We begin by evaluating \deepMergeTool{}'s performance on \emph{synthesizing} resolutions.
%In the resolution synthesis setting, we evaluate \deepMergeTool{} on samples in which $(A, B, O)$ has a true resolution $R$ distinct from $\bot$. 
In this section, we perform an evaluation to assess \deepMergeTool{}'s effectiveness of synthesizing resolutions. Our prediction, $\hat{R}$, is considered correct if it is an exact (line for line, token for token) match with $R$.

\emph{Evaluation metrics.}
\deepMergeTool{} produces a ranked list of predictions; we define top-1 (resp. top-3) accuracy if the $R$ is present in first (resp. top 3) predictions.
This is a lower bound, as multiple resolutions may be ``correct'' with respect to the semantics of the changes being merged (e.g., in some cases, switching two declarations or unrelated statements has no impact on semantics).

\emph{Quantitative Results.}
Table~\ref{tab:full_result_JS} shows the performance of \deepMergeTool{} on a held out test set. 
\deepMergeTool{} has an overall top-1 accuracy of 36.5\%, correctly generating more than one in three resolutions as its first ranked choice. 
When we consider the top-3 ranked resolutions, \deepMergeTool{} achieves a slightly improved accuracy of 43.23\%.
%\shuvendu{@Mayur} We also obtain an upper bound of 43\% for the top-1 predictions that may account for semantically equivalent reordering. 
%\chris{we need details on this.  What does this upper bound mean?}

\begin{table}[t]
\centering
\resizebox{.3\textwidth}{!}{%
\scriptsize
\begin{tabular}{cccc}
\toprule
 %\cmidrule(l){1-3}
 \scriptsize & \scriptsize Top-1 & \scriptsize Top-3  \\
 %\cmidrule(l){1-3}
 \midrule
 {\scriptsize \deepMergeTool{}}& \scriptsize  \textbf{36.50\%} &\scriptsize \textbf{43.23\%} \\
 %\midrule
 \scriptsize \scanMerge{}  & \scriptsize 4.20\% & \scriptsize 7.43\% \\
 %\midrule
 \scriptsize \fairseq  & \scriptsize 2.3\% & \scriptsize 3.3\% \\
 \scriptsize \fstMerge  & \scriptsize 3.7\% & \scriptsize N/A \\
 \bottomrule
\end{tabular}
}
\caption{Evaluation of \deepMergeTool{} and baselines: resolution synthesis accuracy (\%). \label{tab:full_result_JS}}
%\vspace{-0.3in}
\end{table}

\emph{Baselines.}
Table~\ref{tab:full_result_JS} also includes a comparison of \deepMergeTool{} to three baselines.
We compare to a heuristic based approach (\scanMerge{}), an off-the-shelf sequence-to-sequence model (\fairseq{}), and a structured AST based approach (\fstMerge{}). 

Our first baseline \scanMerge{}, is a heuristic based approach designed by manually observing patterns in our dataset. \scanMerge{} randomly samples from the space of sub-sequences over lines from $A$ and $B$ that are:
(i) syntactically valid and parse,
(ii) include each line from $A$ and $B$, and
(iii) preserve the order of lines within $A$ and $B$.
%However, randomly sampling from this space yields 0\% accuracy as 
%the number of such subsequences is exponential. Limiting the space to syntactically valid subsequences only prunes the space close to 50\% on average. Consequently, we further constrain \scanMerge{} to prune trivially incorrect resolutions from the search space.
%(iii) we only allow syntactically valid resolutions that parse when inserted into the entire file. 
%Together, these conditions aid \scanMerge{} by reducing the chance of selecting a trivially incorrect resolution.
These heuristic restrictions are based on manual observations that a large fraction of resolutions satisfy these conditions. 
%These restrictions help limit the set of possible resolutions to randomly select from, to a manageable size. 
%With these three restrictions, the set of possible resolutions to randomly select from shrinks to a manageable size. 

%Consider the example in Figure~\ref{fig:hard}. Although it is a small merge with 4 input lines, without the aforementioned restrictions, there are 64 possible sub-sequences, yielding an accuracy of 1.5\%! By adding restriction (i) we prevent \scanMerge{} from producing trivial resolutions such as \myline{1,B}, \myline{3,B}: \begin{verbatim}               try{ 
%	               } catch (e) {} \end{verbatim} This segment successfully parses in JavaScript suggesting that restricting to syntactically correct resolutions is not sufficient.
%By adding restriction (ii) we prevent \scanMerge{} from producing trivially incorrect, but syntactically correct, resolutions such as the following \myline{2,B}, \myline{1,A}: 
%\begin{verbatim}return require('preconstruct').aliases.webpack(__dirname);
%react-dom = require.resolve('react-dom'); \end{verbatim}

Table~\ref{tab:full_result_JS} shows \scanMerge{}'s performance averaged over 10 trials.
\deepMergeTool{} performs significantly better in terms of top-1 resolution accuracy (36.50\% vs 4.20\%). 
\scanMerge{} only synthesizes one in 20 resolutions correctly. In contrast, \deepMergeTool{} correctly predicts one in 3 resolutions. On inputs of 3 lines or less, \scanMerge{} only achieves 12\% accuracy suggesting that the problem space is large even for small merges.
%Furthermore, \deepMergeTool{} considerably outperforms \scanMerge{} when synthesizing resolutions for inputs with 3 lines or less. \deepMergeTool{} correctly synthesizes over 75\% of resolutions with inputs of 3 lines or less while \scanMerge{} only achieves 12\% accuracy on the same samples. 
%%A further in-depth analysis of \deepMergeTool{}'s effectiveness on merges with varying input sizes is also included in this section.

We also compared \deepMergeTool{} to an out of the box sequence-to-sequence encoder-decoder model~\cite{seq2seq1} (\fairseq) implemented with \textsc{Fairseq}~\footnote{https://github.com/pytorch/fairseq} natural language processing library. Using a \naive{} input (i.e., $concat_3(A, B, O)$), tokenized with a standard byte-pair encoding, and \textsc{Fairseq}'s default parameters, we trained on the same dataset as \deepMergeTool{}. 
%also in \deepMergeTool{}. 
%We trained this model on the same dataset and did not modify the default parameters shipped with fairseq.
%\chris{can we add some details here?  How big is it?  Hyperparameters?  Trained on same data as deepmerge, correct?  Can we provide a link to our code/implementation maybe?}
\deepMergeTool{} outperforms the sequence-to-sequence model in terms of both top-1 (36.5\% vs. 2.3\%) and top-3 accuracy (43.2\% vs. 3.3\%).
This is perhaps not surprising given the precise notion of accuracy that does not tolerate even a single token mismatch.
We therefore also considered a more relaxed measure, the BLEU-4 score~\cite{papineni2002bleu}, a metric that compares two sentences for ``closeness'' using an n-gram model. 
The sequence-to-sequence model achieves a respectable score of 27\%, however \deepMergeTool{} still outperforms with a BLEU-4 score of 50\%.
This demonstrates that our novel embedding of the merge inputs and pointer network style output technique aid \deepMergeTool{} significantly and outperform a state of the art sequence-to-sequence baseline model. 

Lastly, we compared \deepMergeTool{} to \fstMerge \cite{ASE2019}, a recent semistructured AST based approach. \fstMerge{} leverages syntactic information by representing input programs as ASTs. With this format, algorithms are invoked to safely merge nodes and subtrees. Structured approaches do not model semantics and can only safely merge program elements that do not have side effects. Structured approaches have been proven to work well for statically typed languages such as Java~\cite{Apel11, lessenich2015}. However, the benefits of semistructured merge hardly translate to dynamic languages such as JavaScript. JavaScript provides less static information than Java and allows statements (with potential side effects) at the same syntactic level as commutative elements such as function declarations.

As a baseline to compare to \deepMergeTool{}, we ran \fstMerge{} with a timeout of 5 minutes. 
Since \fstMerge{} is a semistructured approach we apply a looser evaluation metric. A resolution is considered correct if it is an exact syntactic match with $R$ \textit{or} if it is semantically equivalent. We determine semantic equivalence manually. \fstMerge{} produces a correct resolution on 3.7\% of samples which is significantly lower than \deepMergeTool{}. Furthermore, \fstMerge{} does not have support for predicting Top-k resolutions and only outputs a single resolution. The remaining 96.3\% of cases failed as follows. In 92.1\% of samples, \fstMerge{} was not able to produce a resolution and reported a conflict. In 3.3\% of samples, \fstMerge{} took greater than 5 minutes to execute and was terminated. In the remaining 0.8\% \fstMerge{} produced a resolution that was both syntactically and semantically different than the user's resolution. In addition to effectiveness, \tool{} is superior to \fstMerge{} in terms of execution time. Performing inference with deep neural approaches is much quicker than (semi) structured approaches. In our experiments, \fstMerge{} had an average execution time of 18 seconds per sample. In contrast, sequence-to-sequence models such as \tool{} perform inference in under a second.

\begin{table}[t]
	\resizebox{.5\textwidth}{!}{%
	\begin{tabular}{cccccc}
	\toprule
		Threshold & [1,3] lines  & [4,5] lines &[6,7] lines & [8,10] lines & [$>$10] lines\\
	 \cmidrule(l){1-6}
	 %\deepMergeTool{}& 76.04\% & 77.88\%  & 46.81\% & 52.66\% & 26.85\% & 35.19\% & 11.96\% & 14.13\% & 1.65\%& 2.20\% \\
	 0 & 78.40\% &  56.50\% &  37.04\% & 10.87\% & 2.93\%\\
	 %.4 & 77.46\% &  52.15\% &  35.19\% & 6.52\% & 1.10\%\\
	 %.5 & 76.06\% & 51.08\% &  30.56\%  & 3.26\% & 0.73\% \\
	 \bottomrule
	\end{tabular}
	}
	%\vspace{-2mm}
	\caption{Evaluation of \deepMergeTool{}: accuracy vs input size\,(\%).\!\!\!\! \label{tab:size_vs_acc}}
	%\vspace{-8mm}
\end{table}

\begin{figure}
	\centering
	%\vspace{-3mm}
	\includegraphics[width=.45\textwidth]{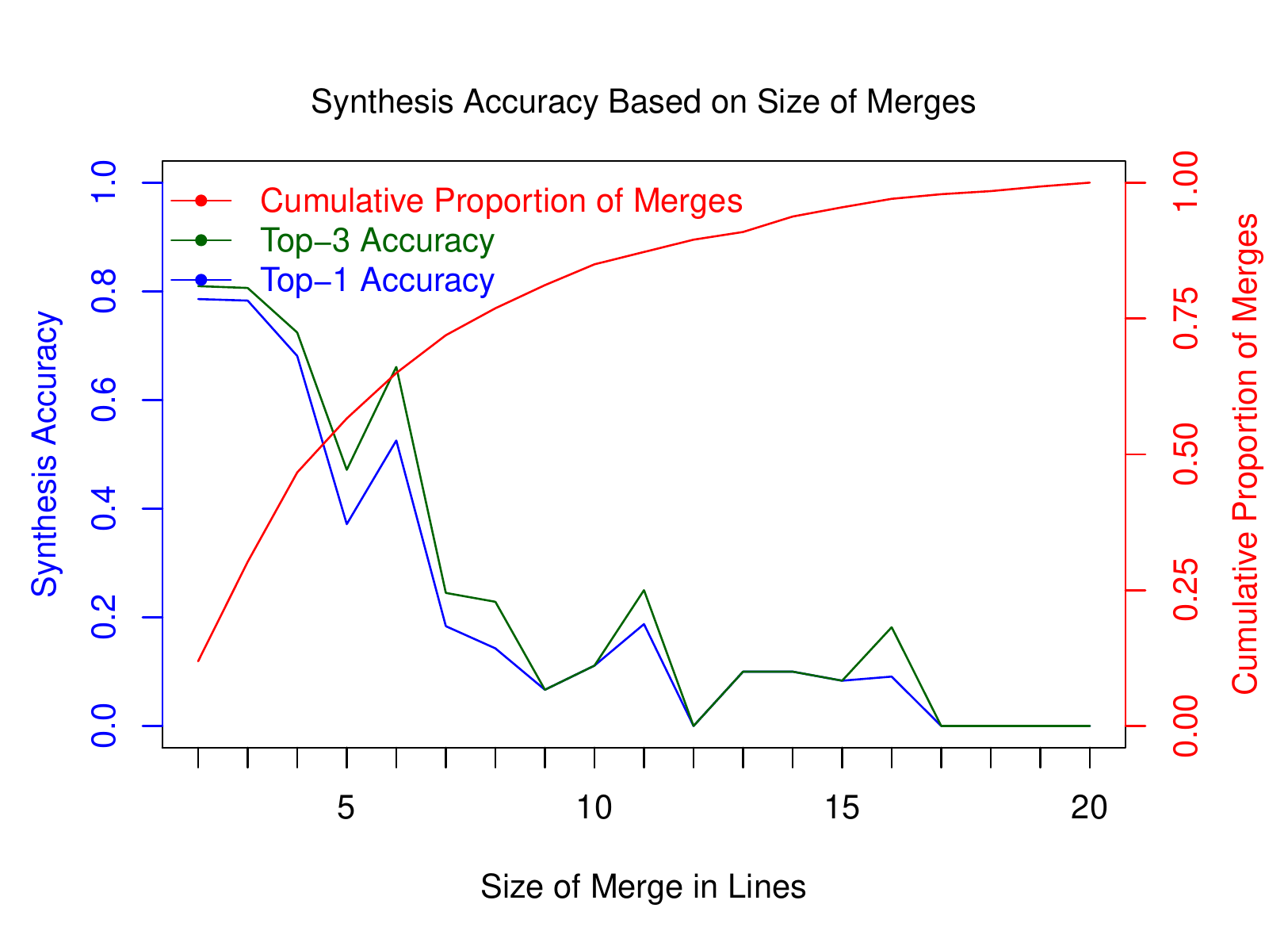}
	%\vspace{-6mm}
\caption{\deepMergeTool{}'s performance vs merge input size. Cumulative distribution of merge sizes in red.\label{fig:size_vs_f1}}
	%\vspace{-4mm}
\end{figure}

%\begin{figure*}[htbp!]
%	\vspace{-0.07in}
%	\begin{subfigure}{0.5\textwidth}
%	\lstset{style=mystyle}
%		\lstinputlisting[linewidth=7.96cm]{./examples/baseline_merged.js}
%		\vspace{-0.1in}
%		\caption{Conflicting merge}
%		\label{fig:a}
%	\end{subfigure}\hfill
%	\begin{subfigure}{0.5\textwidth}
%		\lstset{style=mystyle}
%		\lstinputlisting[linewidth=7.96cm]{./examples/baseline_resolved.js}
%		\vspace{-0.1in}
%		\caption{Merge resolution that \deepMergeTool{} can correctly synthesize}
%		\label{fig:a}
%	\end{subfigure}\hfill
%	\vspace{-3mm}
%	\caption{The set of possible resolutions includes syntactically valid, yet trivially incorrect programs.}
	%\caption{Example of a merge that does not include every line editted from O.}
%	\label{fig:hard}
	%\vspace{-0.07in}
%\end{figure*}

%\begin{comment}
%\begin{figure*}[htbp!]
%	%\vspace{-0.07in}
%	\begin{subfigure}{0.5\textwidth}
%		\lstset{style=mystyle}
%		\lstinputlisting[linewidth=6.96cm]{./examples/oov_merged_trimmed.js}
%		%\vspace{-0.1in}
%		\caption{Conflicting merge}
%		\label{fig:a}
%	\end{subfigure}\hfill
%	\begin{subfigure}{0.5\textwidth}
%		\lstset{style=mystyle}
%		\lstinputlisting[linewidth=6.96cm]{./examples/oov_resolved_trimmed.js}
%		%\vspace{-0.1in}
%		\caption{Merge resolution}
%		\label{fig:a}
%%%%%	\end{subfigure}\hfill
%	%\vspace{-3mm}
%	\caption{$\bot$ Example. A true conflict is detected here as both $A$ and $B$ edit the semantics at the location}
%	\label{fig:oov}
%	%\vspace{-2mm}
%\end{figure*}
%\end{comment}

\emph{Sensitivity to Input Merge Conflict Size.}
We observe that there is a diverse range in the size of merge conflicts (lines in $A$ plus lines in $B$).
However, as shown in Figure~\ref{fig:size_vs_f1}, most (58\% of our test set) merges are small, consisting of 7 or less lines.
As a product of the dataset distribution and problem space size, \deepMergeTool{} performs better for smaller merges.
%The distribution shows that the majority of merges are fairly small, with two and three line merges representing 24\% of all samples with an accuracy of 78.4\%.
%) and two thirds of merges being six lines or less (aggregate accuracy of 67.5\%). 
We present aggregate Top-1 accuracy for the input ranges in Table~\ref{tab:size_vs_acc}. 
\deepMergeTool{} achieves over 78\% synthesis accuracy on merge inputs consisting of 3 lines or less. 
On merge inputs consisting of 7 lines or less (58\% of our test set) \deepMergeTool{} achieves over 61\% synthesis accuracy.

\subsection{RQ2: Effectiveness of Suppressing Incorrect Resolutions}
The probabilistic nature of \deepMergeTool{} allows for accommodating a spectrum of users with different tolerance for incorrect suggestions. ``Confidence'' metrics can be associated with each output sequence to suppress unlikely suggestions. In this section, we study the effectiveness of \deepMergeTool{}'s confidence intervals.

\begin{figure}
	\centering
	%\vspace{-3mm}
	\includegraphics[width=.45\textwidth]{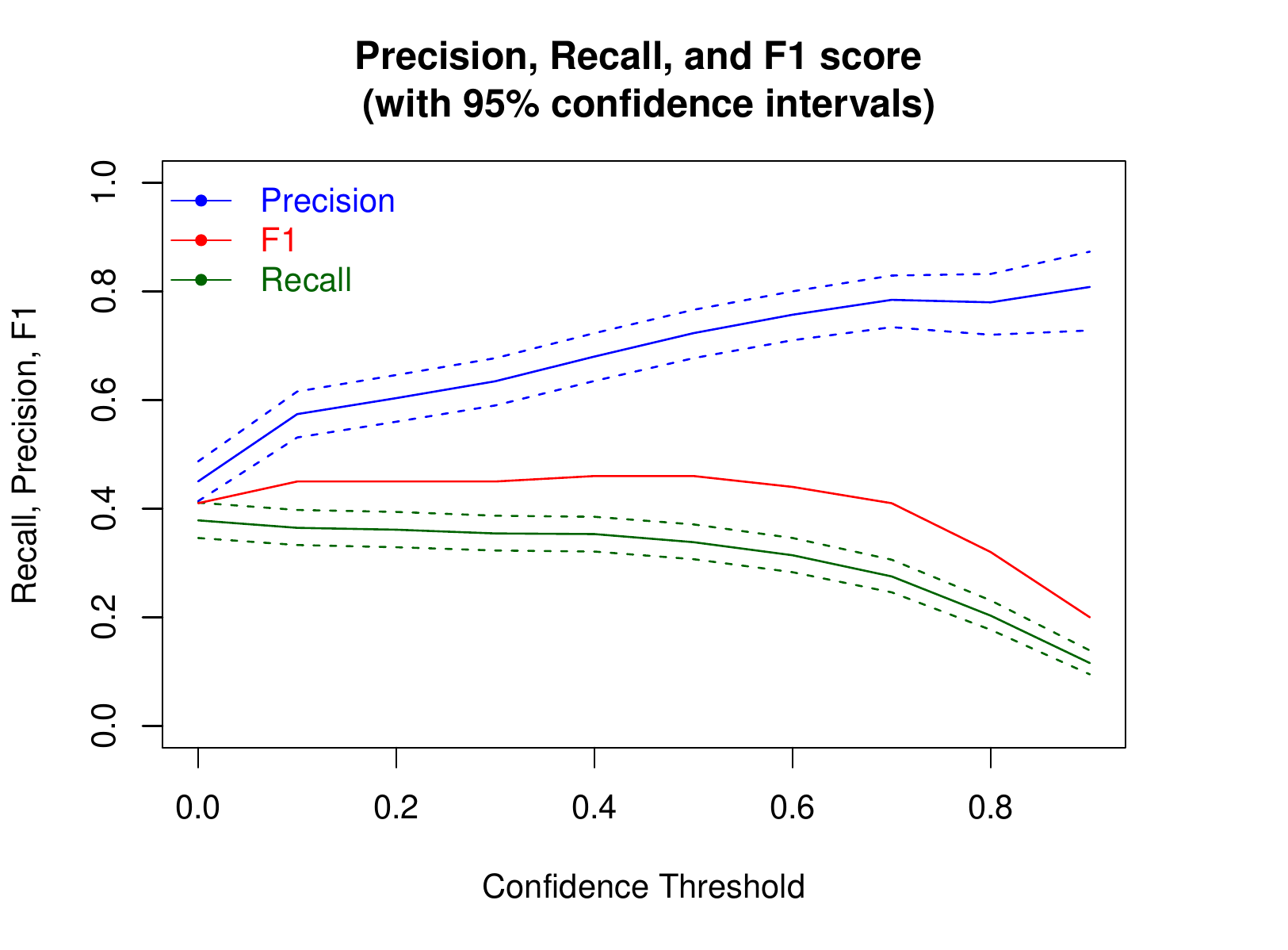}
	%\vspace{-6mm}
	\caption{Top-1 precision and recall by confidence threshold. \label{fig:thresholds}}
	%\vspace{-4mm}
\end{figure}

In the scenario where \deepMergeTool{} cannot confidently synthesize a resolution, it declares a conflict and remains silent without reporting a resolution.  This enables \deepMergeTool{} to provide a higher percentage of correct resolutions (higher precision) at the cost of not providing a resolution for every merge (lower recall). This is critical for practical use, as prior work has shown that tools with a high false positive rate are unlikely to be used by developers~\cite{johnson2013don}.
Figure \ref{fig:thresholds} depicts the precision, recall, and F1 score values, for various confidence thresholds (with 95\% confidence intervals).
We aim to find a threshold that achieves high precision without sacrificing too much recall. 
%The F1-Score is a metric that allows us to compare the performance through both precision and recall with a single value. A threshold that does poorly in either recall \textit{or} precision will have a low F1 score. \TODO{(citation)}
In Figure \ref{fig:thresholds}, the highest F1-Score of 0.46 is achieved at 0.4 and 0.5. 
At threshold of 0.5, \deepMergeTool{}'s top-1 precision is 0.72 with a recall of 0.34.
Thus, while \deepMergeTool{} only produces a resolution one third of the time, that resolution is correct three out of four times. 
Compared to \deepMergeTool{} with no thresholding, at a threshold of .5 \deepMergeTool{} achieves a 2x improvement in precision while only sacrificing a 10\% drop in recall. Thresholds of 0.4 and 0.5 were identified as best performing on a held out validation set. We then confirmed that these thresholds were optimal on the held out test set reported in Figure \ref{fig:thresholds}.

\begin{comment}
\begin{table}[t]
\resizebox{.3\textwidth}{!}{%
\begin{tabular}{cc|c|c}
\toprule
  \multicolumn{1}{c}{Threshold} & \multicolumn{1}{c}{Precision}  & \multicolumn{1}{c}{Recall}  & \multicolumn{1}{c}{F1-Score} \\
 \cmidrule(l){1-4}
 0.0 & .45 & .37 & .41 \\
 0.1 & .57 & .36 & .45 \\
 0.2 & .60 & .36 & .45 \\
 0.3 & .63 & .35 & .45 \\
 \textbf{0.4} & .68 & .35 & \textbf{.46} \\
 \textbf{0.5} & .72 & .34 & \textbf{.46} \\
 0.6 & .76 & .31 & .44 \\
 0.7 & .78 & .28 & .41 \\
 0.8 & .78 & .20 & .32 \\
 0.9 & .81 & .12 & .20 \\
 \bottomrule
\end{tabular}
}
%\vspace{-2mm}
\caption{Precision and Recall tradeoffs via threshold finetuning. \label{tab:threshold_result}}
%\vspace{-0.15in}
\end{table}
\end{comment}

\subsection{RQ3: Categorical Analysis of Effectiveness}
We now provide an analysis of \deepMergeTool{}'s performance. To understand which samples \deepMergeTool{} is most effective at resolving, we classify the dataset into two classes: \concat{} and \other{}. The classes are defined as follows:
\begin{enumerate}[leftmargin=*]
	\item  \textbf{\concat} - resolutions of the form $AB$ or $BA$. Specifically:
		\begin{itemize}
			\item $R$ contains all lines in $A$ and all lines in $B$.
			\item There is no interleaving between $A$'s lines and $B$'s lines.
			\item The order of lines within $A$ and $B$ is preserved.
		\end{itemize}
	\item  \textbf{\other} - resolutions not classified as \concat{}. \\
		\other{} samples can be any interleaving of any subset of lines.
\end{enumerate}

\begin{table}[t]

\centering
\resizebox{.4\textwidth}{!}{%
\begin{tabular}{ccc}
\toprule
	Class & Top-1 Accuracy & Percent of Dataset\\
 \cmidrule(l){1-3}
	\concat{} &  44.40\%  & 26.88\%\\
	\other{} &  29.03\%  & 73.12\%\\
 \bottomrule
\end{tabular}
}
%\vspace{-2mm}
\caption{Accuracy and Distribution of classes. \label{tab:difficulty}}
%\vspace{-8mm}
\end{table}

Table~\ref{tab:difficulty} shows the performance of \deepMergeTool{} on each class. \deepMergeTool{} performs comparably well on each category suggesting that \deepMergeTool{}
is effective at resolving conflicts beyond concatenation.

%\deepMergeTool{} performs best on ?? samples with 50\% accuracy.
%We call these ``remove base'' samples as they take each line that differs from base.  This is a common pattern that our model often correctly predicts.

\subsection{RQ4: Impact of Input Representation}
We now evaluate the use of \textit{Merge2Matrix} and show the benefit of the \textbf{Aligned Linearized} implementation used in \deepMergeTool{}.

We evaluate \deepMergeTool{} on each combination of summarization and edit aware alignment described in Section 3.2: \emph{Na\"ive}, \emph{Linearized}, \emph{LTRE}, \emph{Aligned \naive}, and \emph{Aligned Linearized}.
%We consider both the conflict detection classification (distinguish $R$ from $\bot$) problem and the resolution synthesis problem.
Table~\ref{tab:input_rep} shows the performance of each input representation on detection and synthesis.
%Applying the $linearize_{\theta}$ summarization function achieves a modest boost in synthesis accuracy over the \naive{} \textit{concat} method. (5.63\% for top-1 and 5.86\% for top-3).
The edit-aware input formats: \emph{LTRE}, \emph{Aligned Na\"ive}, and \emph{Aligned Linearized} attain an improvement over the edit-unaware formats. 
%The original \emph{LTRE} representation only produces an improvement over the edit-unaware \emph{Linearized} format (8.12\% for top-1 and 9.26\% for top-3), but slightly worsens detection F1 (-.01).
Our \emph{Aligned} representations are more succinct and contribute to a large increase in accuracy over the edit-unaware formats. 
\emph{Aligned Na\"ive} increases accuracy over our best edit-unaware format by 12.16\% for top-1 and 12.27\% for top-3. 
%F1 score also increases by .01. 
We believe this is due to the verbosity of including the underlying tokens as well as the $\Delta$ edit sequence.
The combination of our edit-aware and summarization insights (\emph{Aligned Linearized}) yields the highest accuracy. % with an increase of  22.55\% top-1 (resp. 21.75\% top-3), and .03  in F1 over its edit-unaware counterpart.

\begin{table}[t]
\centering
\resizebox{.3\textwidth}{!}{%
\begin{tabular}{ccc}
\toprule
 & Top-1 & Top-3 \\
 \cmidrule(l){1-3}
 Na\"ive&  9.62\% & 14.09\%  \\
 Linearized&  15.25\% & 19.95\%  \\
 LTRE&  23.37\% & 29.21\% \\
 Aligned Na\"ive&  27.41\% & 32.22\%  \\
 Aligned Linearized&  \textbf{36.50\%} & \textbf{43.23\%} \\
 \bottomrule
\end{tabular}
}
%\vspace{-2mm}
\caption{Accuracy of different input representation choices. \label{tab:input_rep}}
%\vspace{-8mm}
\end{table}

\subsection{Summary of Results}
%\shuvendu{Compress, too much redundancy with earlier subsections.}
Our evaluation and baselines indicate that the problem of synthesizing resolutions is a non-trivial task, even when restricted to resolutions that rearrange lines from the conflict. 
\deepMergeTool{} not only can synthesize resolutions for more than a third of times, but can also use its internal confidence to achieve high precision (72\%). \deepMergeTool{} can synthesize resolutions significantly more accurately than heuristic based, neural, and structured approaches. We also illustrate the need for edit-aware aligned encoding of merge inputs to help deep learning be more effective synthesizing non-trivial resolutions. 

% !TeX root = ./main.tex
\section{Related Work}
\label{sec:related}
Our technique is related to several existing works in both program merging and deep learning for code.

\subsection{Source Code Merging}
The most widely used method for merging changes is \texttt{diff3}, the default for most version control systems.
One reason for its popularity is that \texttt{diff3} is purely text based and therefore language agnostic. 
However, its behavior has been formalized and Khanna \emph{et al.} showed that the trust developers have in it may be misplaced~\cite{khanna2007formal}, including the examples in Figure~\ref{fig:intro:example}.

There have been many attempts to improve merge algorithms by taking language specific analyses into account (see the work of Mens for a broad survey~\cite{mens2002state}).
Westfechtel \emph{et al.} use the structure of the source code to reduce merge conflicts~\cite{westfechtel1991structure}.
Apel \emph{et al.} noted that structured and unstructured merge each has strengths and weaknesses.  
They developed \textsc{jsFSTMerge}, a \emph{semi-structured merge}, that alternates between approaches~\cite{apel2010semistructured}.
They later introduced \textsc{JDime}, an approach that automatically tunes a mixture of structured and unstructured merge based conflict locations~\cite{apel2012structured}.
Sousa \emph{et al.} introduced a verification approach, \textsc{SafeMerge} that examines the base program, both changed programs, and the merge 
resolution to verify that the resolution preserves semantic conflict freedom~\cite{Sousa18}. 

The key difference between \deepMergeTool{} and these structured or semi-structured merge approaches is that they require a priori knowledge of the language of the merged code in the form of a parser or annotated grammar (or more advanced program verification tools).
Further, structured merge tools cannot conservatively merge changes made within method bodies.
Finally, Pan et al. ~\cite{pan-icse21} explore the use of program synthesis for learning repeated resolutions in a large project.
The approach requires the design of a domain-specific languages inspired by a small class of resolutions (around imports and macros in C++).
In contrast to both these approaches, \deepMergeTool{} only requires a corpus of merge resolutions in the target language, and can apply to all merge conflicts.
However, we believe that both these approaches are complementary and can be incorporated into \deepMergeTool{}.

\subsection{Deep Learning on Source Code}

We leverage deep neural network based natural language processing methods to address the challenge of three way merge resolution.
%There is a wide set of literature on natural language processing and sequence-to-sequence learning and we provide an overview, 
We discuss related works in sequence-to-sequence learning that inspired our model and applications of deep learning for the software engineering domain.
  
% We introduce an extension of seq2seq models to copy lines. Talk about other related works: CopyThat, pointer networks, MSRA work
%\deepMergeTool{} tackles the problem of a very large output vocabulary by using an approach inspired by pointer networks  to copy lines or segments from the input sequence into the output.
%Sequence-to-sequence models introduced by~\citet{seq2seq1,seq2seq2} learn to generate sequences token-by-token, optionally copying tokens from the input sequences. 
Pointer networks~\cite{NIPS2015_29921001} use attention to constrain general sequence-to-sequence models~\cite{seq2seq1,seq2seq2}.
%, which uses attention for the same purpose.  
Recent works incorporate a {\it copy} mechanism in sequence-to-sequence models by combining copying and token generation~\cite{gu-etal-2016-incorporating}, adding a copying module in the decoder~\cite{Zhou2018SequentialCN}, and incorporating it into the beam search~\cite{panthaplackel2020copy}.
%Gu \textit{et al.}\citet{gu-etal-2016-incorporating} who modeled translation as a combination of token generation and copying, 
%\citet{Zhou2018SequentialCN} who incorporated a copying module into the model decoder, and~\citet{panthaplackel2020copy} who modified the beam search algorithm to facilitate span-copying.
%
In contrast to \tool, none of these approaches address the challenges described in Section~\ref{sec:formulation} in a three-way merge. 
%None of these approaches, however, fully leverage the line-level nature of the three-way textual program merge in their model architecture. 
%In this paper, we impose stronger constraints on the output vocabulary to better fit our task, as described in Section~\ref{sec:output}.

% NLP for automated software engineering in general
Deep learning has been successfully used on source code to improve myriad software engineering tasks. 
These include code completion and code generation~\cite{10.1145/3292500.3330699,clement2020pymt5}, code search~\cite{codeSearch2018}, 
software testing~\cite{godefroid2017learn}, defect prediction~\cite{wang2016automatically}, and code summarization~\cite{alon2018codeseq}.
Deep learning has been used in program repair using neural machine translation~\cite{10.1145/3340544,DBLP:journals/corr/abs-1810-00314}, sequence-editing approaches~\cite{panthaplackel2020copy}, 
and learning graph transformations~\cite{hoppity}.
For a deeper review of deep learning methods applied to software engineering tasks, see the literature reviews\cite{li2018deep, ferreira2019software}.

While neural sequence-to-sequence models are utilized in most of those applications, they consume only one input sequence, mapping it to a single output sequence. 
Edit aware embeddings~\citet{yin2018learning} introduced LTRE method to encode two program variants to model source code edits.
As we demonstrate, our edit-aware encoding \textbf{Aligned Linearized} is inspired by this approach but significantly outperforms LTRE in the context of data-driven merge.
%in our approach to the three-way program merge setting (see Section \ref{sec:deepmerge}) and \mergetomatrix is the first to incorporate three program variants as neural model inputs. 

% !TeX root = ./main.tex
\section{Conclusion}
\label{sec:conclusion}

We motivated the problem of {\it data-driven merge} and highlighted the
main challenges in applying machine learning.
We proposed \deepMergeTool{}, a data-driven merge framework, and
demonstrated its effectiveness in resolving unstructured merge conflicts in
JavaScript.
We chose JavaScript as the language of focus in this paper due to its importance and growing popularity and the fact that analysis of JavaScript is challenging due at least in part to its weak, dynamic type system and permissive nature~\cite{jensen2009type,kashyap2014jsai}.
We believe that \deepMergeTool{} can be easily extended to other languages
and perhaps to any list-structured data format such as JSON and configuration files. 
We plan to combine program analysis
techniques (e.g., parsing, typechecking, or static verifiers for merges)
to prune the space of resolutions, and combine structured
merge algorithms with machine learning to gain the best of both techniques.
Furthermore, we plan to generalize our approach beyond line level output granularity. 
%While we have intentionally treated code as line-structured data and have not introduced any deeper semantic analysis in an effort to enable \deepMergeTool{} to be as general as possible, we have not trained it or evaluated it on other languages.
%Thus, while we are optimistic that it will work for other languages and no changes to \deepMergeTool{} are required to for that, we cannot claim that it's use would yield similar results for languages other than JavaScript.

%reference
\bibliographystyle{abbrv}
\bibliography{refs}

\end{document}